\documentclass[5p,number]{elsarticle}

\usepackage{amsmath}
\usepackage{graphicx}
\usepackage{float}
\usepackage{textcomp}	
\usepackage{dcolumn}
\usepackage{hyperref}
\renewcommand{\epsilon}{\varepsilon}
\journal{Energy Policy}

\begin{document}
\begin{frontmatter}
\title{Historical Costs of Coal-Fired Electricity and Implications for the Future}
\author[BU,SFI]{James McNerney}
\author[SFI,LUISS]{J. Doyne Farmer}
\author[MIT,SFI]{Jessika E Trancik\corref{cor}}
\cortext[cor]{Corresponding author. \emph{E-mail address:} trancik@mit.edu}
\address[BU]{Department of Physics, Boston University, Boston, MA 02215, USA}
\address[SFI]{Santa Fe Institute, 1399 Hyde Park Road, Santa Fe, NM 87501, USA}
\address[LUISS]{LUISS Guido Carli, Viale Pola 12, 00198, Roma, Italy}
\address[MIT]{Engineering Systems Division, Massachusetts Institute of Technology, Cambridge, MA 02139-4307, USA}

\begin{abstract}
We study the costs of coal-fired electricity in the United States between 1882 and 2006 by decomposing it in terms of the price of coal, transportation costs, energy density, thermal efficiency, plant construction cost, interest rate, capacity factor, and operations and maintenance cost.  The dominant determinants of costs have been the price of coal and plant construction cost.   The price of coal appears to fluctuate more or less randomly while the construction cost follows long-term trends, decreasing from 1902 - 1970, increasing from 1970 - 1990, and leveling off since then. Our analysis emphasizes the importance of using long time series and comparing electricity generation technologies using decomposed total costs, rather than costs of single components like capital. By taking this approach we find that the history of coal-fired electricity costs suggests there is a fluctuating floor to its future costs, which is determined by coal prices. Even if construction costs resumed a decreasing trend, the cost of coal-based electricity would drop for a while but eventually be determined by the price of coal, which fluctuates while showing no long-term trend.
\end{abstract}

\begin{keyword}
coal \sep electricity \sep historical cost
\end{keyword}
\end{frontmatter}

\tableofcontents

\section{Introduction}
Coal generates two-fifths of the world's electricity \cite{WDI} and almost a quarter of its carbon dioxide emissions \cite{IEA08}. 
The impact of any market-based effort to reduce carbon emissions will be highly sensitive to future costs of coal-fired electricity in comparison to other energy technologies.  The relative cost of technologies will, for example, determine the carbon reductions resulting from a particular carbon tax or the cost of a given cap on emissions. However, no study exists that examines total generation costs and component contributions using data over a time span comparable to that of the forecasts needed. This study attempts to fulfill this need. We also aim to make methodological advances in the analysis of historical energy costs and implications for future costs. 

To do this, we build a physically-accurate model of the total generation cost in terms of the price of coal, coal transportation cost, coal energy density, thermal efficiency, plant construction cost, interest rate, capacity factor, and operations and maintenance cost. This contrasts with the approach taken in econometrics where generation costs, or more typically plant costs, are broken down using a regression model \cite{Komiya62,Nerlove63,Cowing74,Christensen76,Huettner78,Stewart79,Nelson83,Gollop83,Joskow85,McCabe96,Hisnanick99}. 

We also focus on the long term evolution of quantities averaged across plants in the United States, going back to the earliest coal-fired power plant in 1882 through 2006, rather than cross-sections or panels of plants. Thus, our data set sacrifices cross-sectional richness to examine a longer time span (over a century). This is important for characterizing the factors driving the long term evolution of costs.

The data suggest a qualitative difference between the behavior of fuel and capital costs, the two most significant contributors to total cost. Coal prices have fluctuated and shown no overall trend up or down; they became the most important determinant of fuel costs when average thermal efficiencies ceased improving in the U.S. during the 1960s.  This fluctuation and lack of trend are consistent with the fact that coal is a traded commodity and therefore it should not be possible to make easy arbitrage profits by trading it.  According to standard results in the theory of finance, this implies that it should follow a random walk.  In contrast, plant construction costs, the most important determinant of capital costs, followed a decreasing trajectory until 1970, consistent with what one expects from a technology. After 1970, construction costs dramatically reversed direction, at least in part due to pollution controls.  

Analysis of historical trends suggests a fluctuating floor on the total costs of coal-fired electricity which is determined by coal prices. Under a scenario in which plant costs return to their pre-1970 rate of decrease, fuel costs would rise in their relative contribution to the total. This would lead to higher uncertainty in total generation costs, and create a floor below which the generation cost would be unlikely to drop. Even under a scenario in which carbon capture and storage (CCS) capabilities are added to plants, the same qualitative behavior would be expected.

Our analysis makes several methodological advances. It calculates total costs of generation, rather than the costs of individual components like capital, and over a long ($\sim$100 year) time span. We build on an approach developed for photovoltaics and nuclear fission \cite{Nemet06,Koomey07} to decompose changes in cost. The refinement eliminates artificial residuals arising from the cross-effects of variables influencing the cost. Physically accurate models like the one presented here may sometimes allow (data permitting) for more reliable decompositions of cost than regression models.

The paper is organized as follows: in Section \ref{historical}, we present the historical record of costs for coal-fired electricity, as represented by time series for a number of important variables influencing generation cost. The generation cost consists of three main \emph{components} --- fuel, capital, and operation and maintenance. The first two are further decomposed; the fuel component into the coal price, transportation cost, coal energy density, and thermal efficiency; the capital component into plant construction costs, capacity factor, and interest rate. In Section \ref{trends}, we determine the variables contributing most to the historical changes in generation costs, focusing on their long term trends rather than their short term variation. In Section \ref{variation}, we examine the effect their short term variation has on total generation costs. In Section \ref{extrapolation}, we use the experience gained from analyzing the data to examine future implications.

\section{Historical data\label{historical}}

\subsection{Data sources}
Sources for data are given in the captions of the relevant figures. The data comes from the U.S. Energy Information Administration, Census Bureau, Bureau of Mines, Federal Energy Regulatory Commission (formerly the Federal Power Commission), Federal Reserve, the Edison Electric Institute, and the Platts UDI world electric power plants database, with some minor contributions from a few other databases and technical reports. Assumptions made to fill in for missing data are noted in the relevant figures and text. 

All data presented here are for the United States between 1882 (the approximate beginning of the electricity utility industry in the U.S.) and 2006. All data are estimates for averages over U.S. coal utility plants. All prices and costs are presented in real 2006 currency, deflated using the GDP deflator.%
\footnote{Given the existence of other industry-specific deflators (e.g. PPIs, the Handy-Whitman indices), it is worth explaining why we deflate all prices by the GDP deflator. To study how the total cost of electricity is affected by the direct inputs to electricity production, later given in equation \eqref{main_equation} --- O\&M, the coal price, transportation price, and the construction price --- we deflate prices in a way that preserves their ratios, while removing the effect of changes in the overall price level of the economy. Such ratios represent a meaningful quantity --- the relative economic scarcity of two goods --- and the real prices should preserve them. Using a single deflator for all these inputs guarantees that their price ratios are unchanged. To remove changes in the overall price level of an economy, the GDP deflator is appropriate. }
The data is made available at www.santafe.edu/files/coal\_electricity\_data.

\subsection{Decomposition formula}
We seek to build up the total generation cost (TC) from the following variables,%
\footnote{By total generation cost, we mean all production costs of electricity up to the busbar, the point at which electricity leaves the plant and enters the grid.}
 for which data is available:
\begin{align*}
\text{OM}   &= \text{total operation and maintenance cost(\textcent/kWh)}\\
\text{FUEL}   &= \text{total fuel cost (\textcent/kWh)}\\
\text{CAP}   &= \text{total capital cost (\textcent/kWh)}.
\end{align*}
These three major cost components are in turn decomposed further into
\begin{align*}
\text{COAL}  &= \text{price of coal (\$/ton)}\\
\text{TRANS} &= \text{price of transporting coal to plant (\$/ton)}\\
\rho  &= \text{energy density of coal (Btu/lb)}\\
\eta  &= \text{plant efficiency}\\
\text{SC}   &= \text{specific construction cost (\$/kW)}\\
r     &= \text{nominal interest rate}\\
\text{CF}    &= \text{capacity factor}.
\end{align*}
\emph{Specific construction cost} (SC) here means the construction cost of a plant per kilowatt of capacity.  Each variable is given a subscript $t$ to denote its value in year $t$. The total generation cost (TC) in year $t$ is
\begin{align} \label{main_equation}
\text{TC}_t &= \text{OM}_t + \qquad \quad \text{FUEL}_t \qquad \ \,+ \qquad \quad \text{CAP}_t\\
&= \text{OM}_t + \frac{\text{COAL}_t + \text{TRANS}_t}{\rho_t \eta_t} + \frac{\text{SC}_t \times \text{CRF}(r_t,n)}{\text{CF}_t \times 8760 \text{ hrs}} \nonumber.
\end{align}
The three main terms are the three major cost components -- operation and maintenance, fuel, and capital. The fuel component accounts for the cost of coal and its delivery to the plant, the amount of energy contained in coal, and the efficiency of the plant in converting stored chemical energy to electricity. The capital component levelizes the construction cost of the plant using the capital recovery factor, $\text{CRF}(r_t,n)$, defined as
\begin{align} \label{crf}
\text{CRF}(r_t,n) = \frac{r_t(1+r_t)^n}{(1+r_t)^n-1},
\end{align}
where $n$ is the plant lifetime in years. The capital recovery factor is the fraction of a loan that must be payed back annually, assuming a stream of equal payments over $n$ years and an annual interest rate $r_t$. For a plant of capacity $K$, the capital component is the annuity payment on money borrowed for construction, $K \times \text{SC}_t \times \text{CRF}(r_t,n)$, divided by the yearly electricity production of the plant, $K \times \text{CF}_t \times 8760 \text{ hrs}$. Note that $K$ cancels out. Thus, the capital component is the annuity payment on borrowed plant construction funds \emph{per} kilowatt-hour of annual electricity production by the plant. We follow convention and levelize over an assumed plant lifetime of $n=30$ years.
 
Note that the total cost for year $t$ uses the capital cost of plants built that year, while the actual fleet of plants existing in year $t$ also includes plants built in previous years. Since we do not have data on the retirement dates of plants, we use the cost of the plants built in year $t$ to estimate the cost of electricity generated in that year. Average capital costs of the whole fleet existing in a particular year have the same basic evolution as that of new plants alone, but tend to lag the latter, since they include capital costs from previous years.

In Sections \ref{fuelcost} -- \ref{capcost}, we present and discuss the time-series data for each of these variables separately. In Section \ref{totalcost} we combine the variables to compute the total cost, and examine the effect that individual variables had on it. The ensuing sections are an extensive discussion of the data; readers more interested in final analyses may wish to skip to Section \ref{trends}.

\subsection{Fuel cost \label{fuelcost}}

\subsubsection{Coal price}
\begin{figure}[t]
\includegraphics[width=0.5\textwidth]{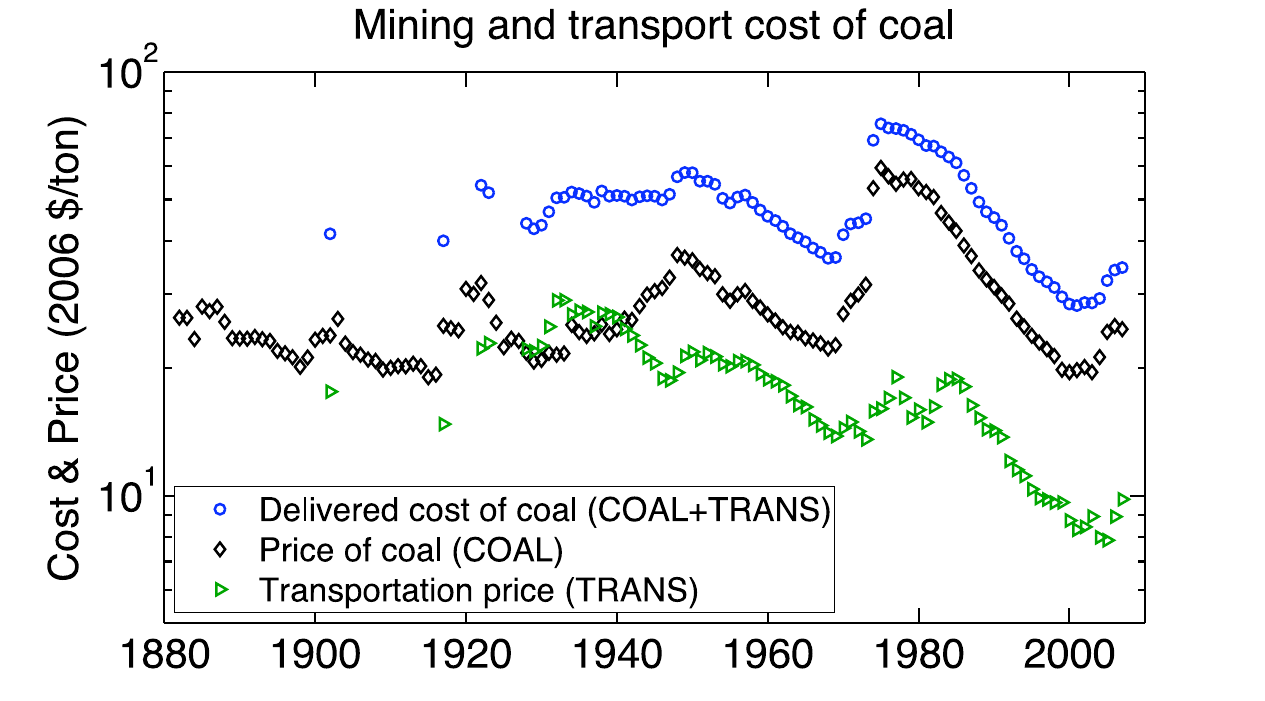}
\caption{The price of coal at the mine, the price of transporting it to the plant, and their sum. The price of coal has fluctuated with no clear trend up or down. Transportation contributed a significant while decreasing expense to fuel costs. Source: \cite{Manthy78,AER06,Schurr60,MineralYearbook,EEI92}}
\label{coalprice}
\end{figure}

Real coal prices (COAL) have varied over the past 130 years (Fig. \ref{coalprice}.) The largest change occurred between 1973 and 1974. A government study \cite{Council76} found this was due to the OPEC oil embargo started in December 1973, which raised prices for substitutes coal and natural gas, and anticipation of a strike by the United Mine Workers union, which later occurred at the end of 1974. Wage increases starting in 1970 also played a somewhat less important role.

Coal prices at the mine were derived from price time-series for individual varieties of coal. For the period 1882-1956, the coal price is a production-weighted average over anthracite and bituminous varieties, and for 1957-2006 over anthracite, bituminous, subbituminous, and lignite. No data could be found to weight prices by the quantities in which they were consumed by plants, and therefore we relied on the quantities in which they were produced. The resulting average price series closely resembles that of bituminous coal by itself.

\subsubsection{Transportation costs}
Transportation costs (TRANS) were derived mainly from the cost of coal delivered to power plants and the price of coal at the mine, though some direct data also exists. Transportation costs have added a significant though decreasing amount to the cost of coal delivered to power plants (Fig. \ref{coalprice}). Transportation costs before 1940 were on par with the price of coal at the mine; since 1940, they have dropped to 20 -- 40\% of the delivered cost on average, though there is considerable variation from plant to plant. The introduction of unit trains in the 1950s may account for some of the decrease in transportation costs. A unit train carries a single commodity from one origin to one destination, shortening travel times and eliminating the confusion of separating cars headed for different destinations. About 50\% of coal shipped in the United States (90\% of which is used by utilities) is carried by unit trains \cite{BritannicaOnline}. Other means of transporting coal are barges, collier ships, trucks, and conveyor belts in cases where plants are built next to mines. Although we focus on average transportation costs, we note that the local cost of coal varies considerably between regions.

Several acts during the 1970s partially deregulated the U.S. rail industry, culminating in the 1980 Staggers Rail Act which completed deregulation \cite{Martland99}. Rail rates before 1980 were determined by the Interstate Commerce Commission; after the Staggers Act, railroads were allowed to determine their own rates. The deregulation is believed to have contributed to lower rates in the following years \cite{Martland99}.

Many utilities responded to environmental regulations by switching to higher priced, low-sulfur coals \cite{Gollop83}. Switching to low sulfur coals also extended transportation distances, which may explain the increase in transportation costs seen around 1970. The 1990 Clean Air Act Amendments tightened regulations again, extending transportation distances \cite{EIACoal2000}. However, a coincident decrease in rail rates per ton-mile caused rates per ton to continue decreasing.

\subsubsection{Coal energy density}
\begin{figure}[t]
\includegraphics[width=0.5\textwidth]{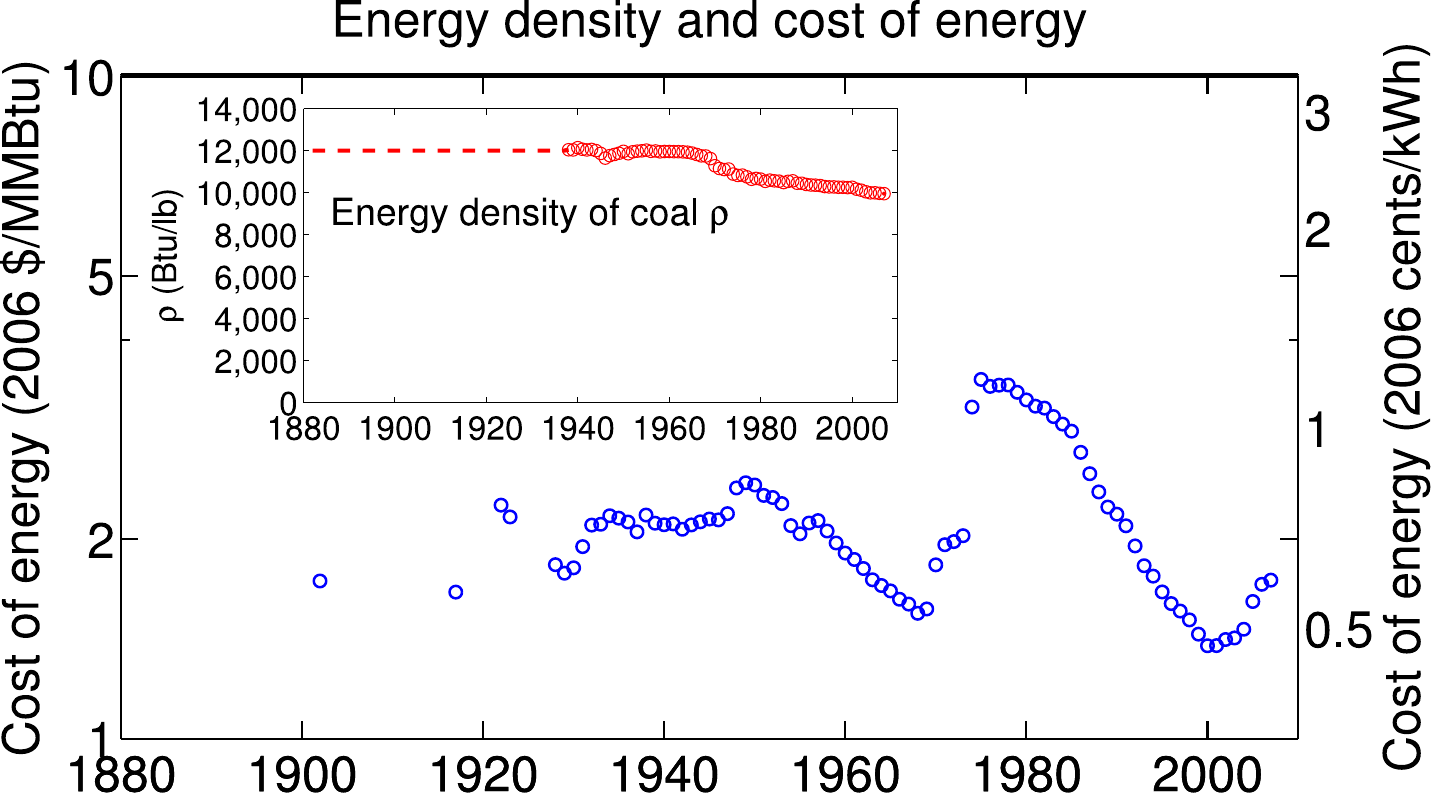}
\caption{The cost of coal-stored energy, $(\text{COAL}_t + \text{TRANS}_t)/\rho_t$. The left axes show cost in dollars per million BTU, the right axes in cents per kWh. The inset shows the energy density of coal in Btu per pound. The dashed line indicates an assumed value of 12,000 Btu per pound. Coal energy densities have come down about 16\% in the last 40 years. Source: \cite{FPC47,AER06}}%
\label{energyprice_density}
\end{figure}

Since 1960, the average energy density of coal ($\rho$) has dropped steadily (Fig. \ref{energyprice_density} inset.) The lower the energy density, the more coal needed by plants to consume equal amounts of primary energy. Thus, the effect of lower energy densities has been to increase fuel costs through increased purchase and transportation costs.

Changes in the energy density reflect changes in the overall mixture of coal species used by the industry, as well as variation of energy density within species. One cause of the decrease in energy density may be the increased use of subbituminous coal, which burns more cleanly than previously used bituminous due to a lower sulfur content, but also contains less energy per pound. No data for energy density could be found before 1938, and a value of 12,000 Btu per pound was assumed (the 1938 value).

\subsubsection{Thermal efficiency} \label{efficiency}
\begin{figure}[t]
\includegraphics[width=0.5\textwidth]{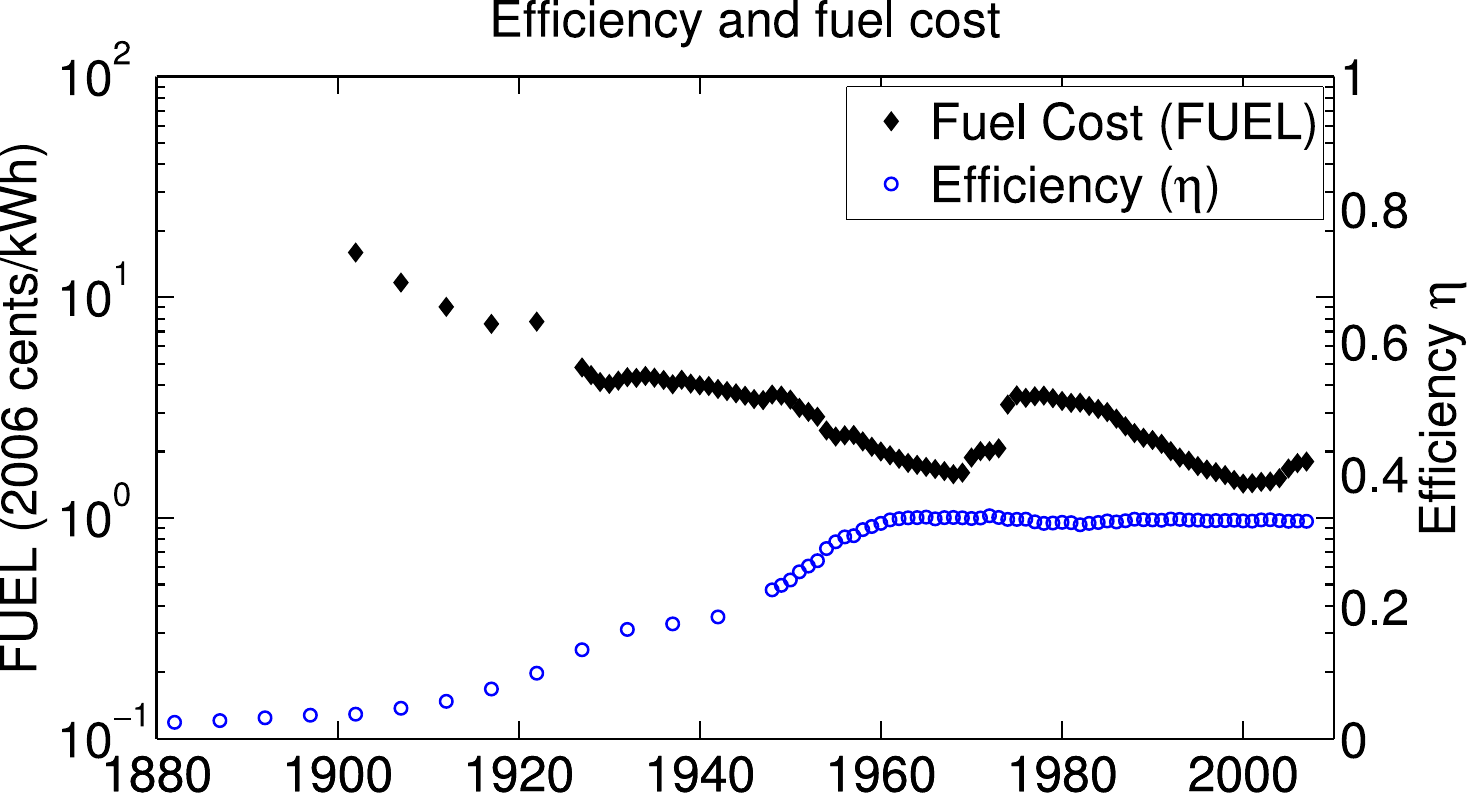}
\caption{Average efficiency of coal plants and cost of the fuel component, $(\text{COAL}_t + \text{TRANS}_t)/(\rho_t \eta_t)$. Efficiency increases drove decreases in fuel costs until 1960. Since then efficiency has remained stable around 32 -- 34\%. Source: \cite{Neil42,AER06} }%
\label{efficiency_fuelcost}
\end{figure}

The thermal efficiency of a fossil-steam plant ($\eta$) is the fraction of stored chemical energy in fuel that is converted into electrical energy (Fig. \ref{efficiency_fuelcost}). It accounts for every effect which causes energy losses between the input of fuel and the bus-bar of the plant, the point where electricity enters the electric grid. These effects include incomplete burning of fuel, radiation and conduction losses, stack losses, excess entropy produced in the turbine, friction and wind resistance in the turbine, electrical resistance, power to run the boiler feed pump and other plant equipment, and the thermodynamic limits of the heat cycle used.

The earliest coal plants obtained efficiencies below 3\%. Over the first 80 years, average efficiency grew by more than a factor of 10.  Improving efficiencies meant that less coal was required to produce equal amounts of electricity, lowering fuel costs. For the last 50 years, the average efficiency of coal-fired plants has stayed approximately level at 32--34\%, although individual plants obtained efficiencies as high as 40\% during the 1960s \cite{Hirsh89}.

\begin{table}[b]
\caption{Operational sources of energy loss. Source: \cite{Hirsh89}}
\label{efficiency_table}
\begin{tabular*}{.48\textwidth}{@{\extracolsep{\fill}}lr}
\hline
\textbf{Source of energy loss} & \textbf{Efficiency}\\
\hline
Stack losses, radiation\\
\quad \& conduction from boiler & 87\%\\
Excess entropy produced in turbine	& 92\%\\
Windage, friction, \& elec. resistance & 95\%\\
Boiler feed pump power requirement &95\%\\
Auxiliary power requirements & 97\%\\
\hline
Total (product)	& 70\%\\
\hline
\end{tabular*}
\end{table}

The factors influencing efficiency of plants fall into two major categories: thermodynamic factors, which 
pertain to the specific heat cycle employed, and operational factors, which reflect the mechanical and electrical efficiencies of individual stages and components. The thermodynamic limit efficiency is determined by the details of the steam cycle used -- the maximum and minimum temperatures, the maximum and minimum pressures, and the exact type of cycle followed (e.g. the number of reheat stages used.) Current plant designs typically have thermodynamic limit efficiencies around 46\%.\footnote{We assume a Rankine cycle with superheating and one reheat cycle operating at 1000 $^\circ$F, 2400 psi.} The operational efficiencies of the boiler, turbine, generator, and other components further reduce the efficiency achievable in practice (Table \ref{efficiency_table}.) The net effect of the operational efficiencies is to reduce total efficiency by a factor of about 0.7.

The capacity factor affects the operating parameters of the plant (e.g. pressures, throttles) and therefore influences plant efficiency. Typically, a lower capacity factors will mean higher variability and lower efficiencies. The presence of pollution controls can also lower efficiency, because they increase auxiliary power requirements.

Historical increases in efficiency came from improvements in both the thermodynamic and operational categories. Higher steam temperatures, higher steam pressures, and changes to the heat cycle relaxed the thermodynamic constraint, while better designed parts reduced operational losses. However, operational factors are now working at high efficiencies (Table \ref{efficiency_table}), while changing the parameters of the heat cycle faces highly diminishing returns for the thermodynamic limit efficiency \cite{Hirsh89}. This latter problem reduces incentives to develop economical materials that could withstand higher pressures and temperatures. The overall result has been a standstill in average efficiency for the last 50 years in the U.S. (Although the average efficiency of U.S. plants has been static in the last 50 years, that of European and Japanese plants has continued to grow in the same period.) Given the present technological options, the current average efficiency presumably represents an optimum after balancing the higher fuel cost of a less efficient plant against the higher construction costs of a more efficient plant.

\subsubsection{Fuel cost}
Over the entire time period (1882-2006) the factors most responsible for changes in the fuel cost are the price of coal and the efficiency. Changes in energy density and transportation costs had relatively minor effects. As can be seen in Fig. \ref{efficiency_fuelcost}, the net effect of varying coal price, energy density, transportation cost, and efficiency on the fuel component of electricity costs was a long term decrease in the cost of the fuel component until about 1970. The decrease was mainly due to improving efficiency. After 1970, coal prices increased dramatically during a time when efficiency was flat, increasing overall fuel costs 70\% between 1970 and 1974.

\subsection{Operation and maintenance cost \label{sec:om}}
The O\&M cost (OM) is the least significant of the three cost components (fuel, O\&M, capital), representing about 5-15\% of total generation costs during the last century (Fig. \ref{component_shares}). Although it is less significant than fuel or capital, we attempted to reconstruct a time-series for O\&M.
\begin{figure}[t]
\includegraphics[width=0.5\textwidth]{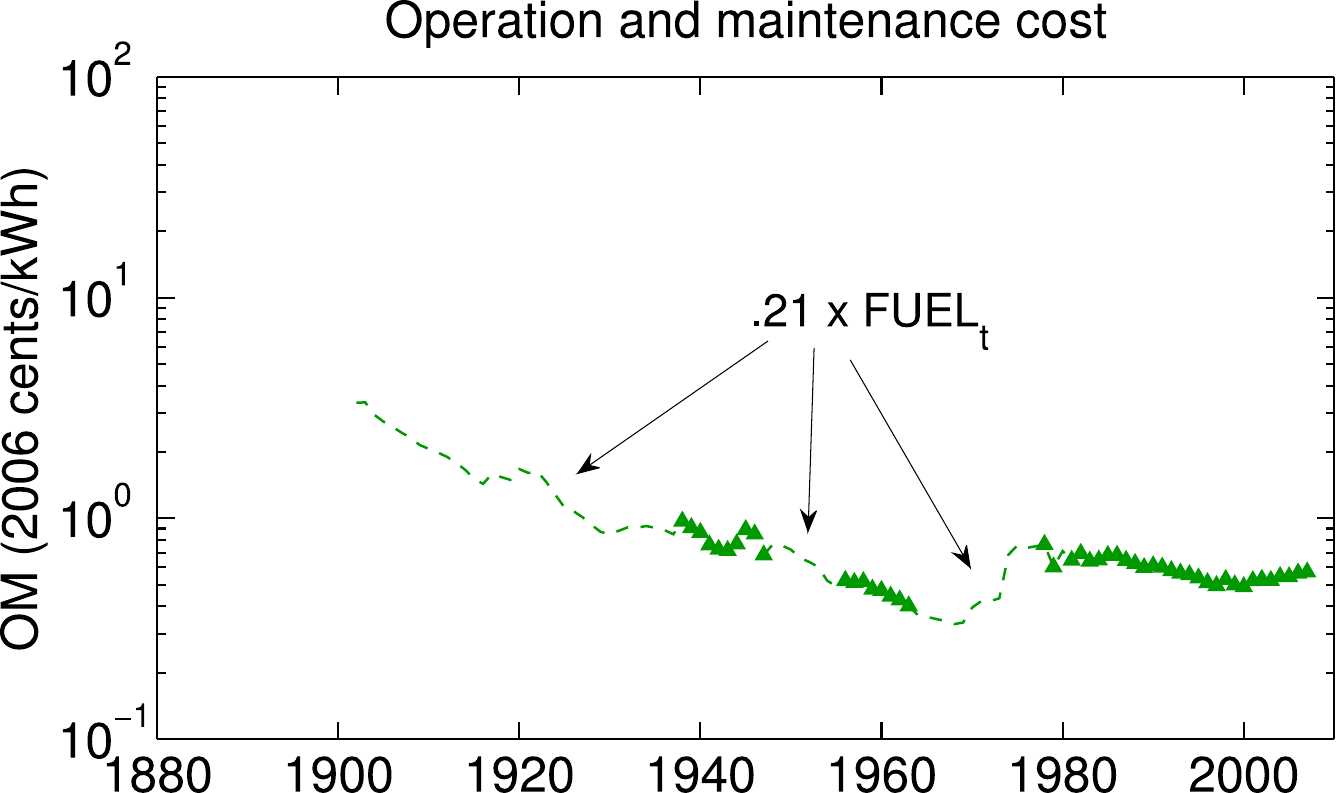}
\caption{The cost of operation and maintenance of coal plants. A value equal to 21\% of the overall fuel cost for a given year was used to fill in missing data. Source: \cite{FPC47,FPC61,EIA78,EIA79,Beamon99,NEI} \label{om}}
\end{figure}

Reliable historical data for O\&M costs was difficult to acquire. Sources were frequently in conflict with each other due to differences in definition of O\&M costs. We use data from the Energy Information Administration (EIA) because it uses a consistent definition over the longest time period. To fill in missing years, we took O\&M costs to be 21\% of the value of the overall fuel cost, based on the empirical observation that O\&M costs were consistently about 21\% of overall fuel costs between 1938 and 1985 in all years for which we had data.%
\footnote{OM's ratio to FUEL was always between 19 and 24\%, with an average of 21\%, during three periods for which data was available: 1938-1947, 1956-1963, and 1978-1985.}%
Later O\&M costs break from this pattern. This assumption seemed the best choice to avoid introducing discontinuities or other artificial behaviors in the total costs. We do not have a theoretical explanation for why the "21\% rule" replicates the O\&M costs so well for the years where data is available. However we note that based on empirical analysis, the O\&M costs are more strongly correlated with the fuel costs than with capital or construction costs.%
\footnote{The correlation coefficient between OM and FUEL is 0.87, while that between OM and CAP is 0.69 and that between OM and SC is 0.33.}

Based on our reconstruction, O\&M costs decreased until 1970 (Fig. \ref{om}). For pulverized coal plants, historical declines ``are attributed mainly to the introduction of single-boiler designs, automatic controls, and improved instrumentation'' \cite{Yeh07,Sporn68}. From the data available, we infer that an increase occurred between 1960 and 1980. This is supported by observations that pollution controls introduced to plants during this time raised O\&M costs \cite{Komanoff81,Joskow85}.

O\&M costs are also influenced by the capacity factor. Higher production rates lower O\&M costs by amortizing cost over greater electricity production \cite{Komanoff81}.

\subsection{Capital cost \label{capcost}}

\subsubsection{Specific construction cost \label{specificCost}}
Construction cost data were obtained from \emph{Historical Plant Cost and Annual Production Expenses 1982} \cite{EIA82}, published by the Energy Information Administration. A potential source of bias with this data is that plant costs were given as accounts which accumulate a utility's nominal expenditure on a given plant. These accounts therefore include costs of additional units since its construction, and sum together nominal expenditures from different years.
\begin{figure}[t]
\includegraphics[width=0.5\textwidth]{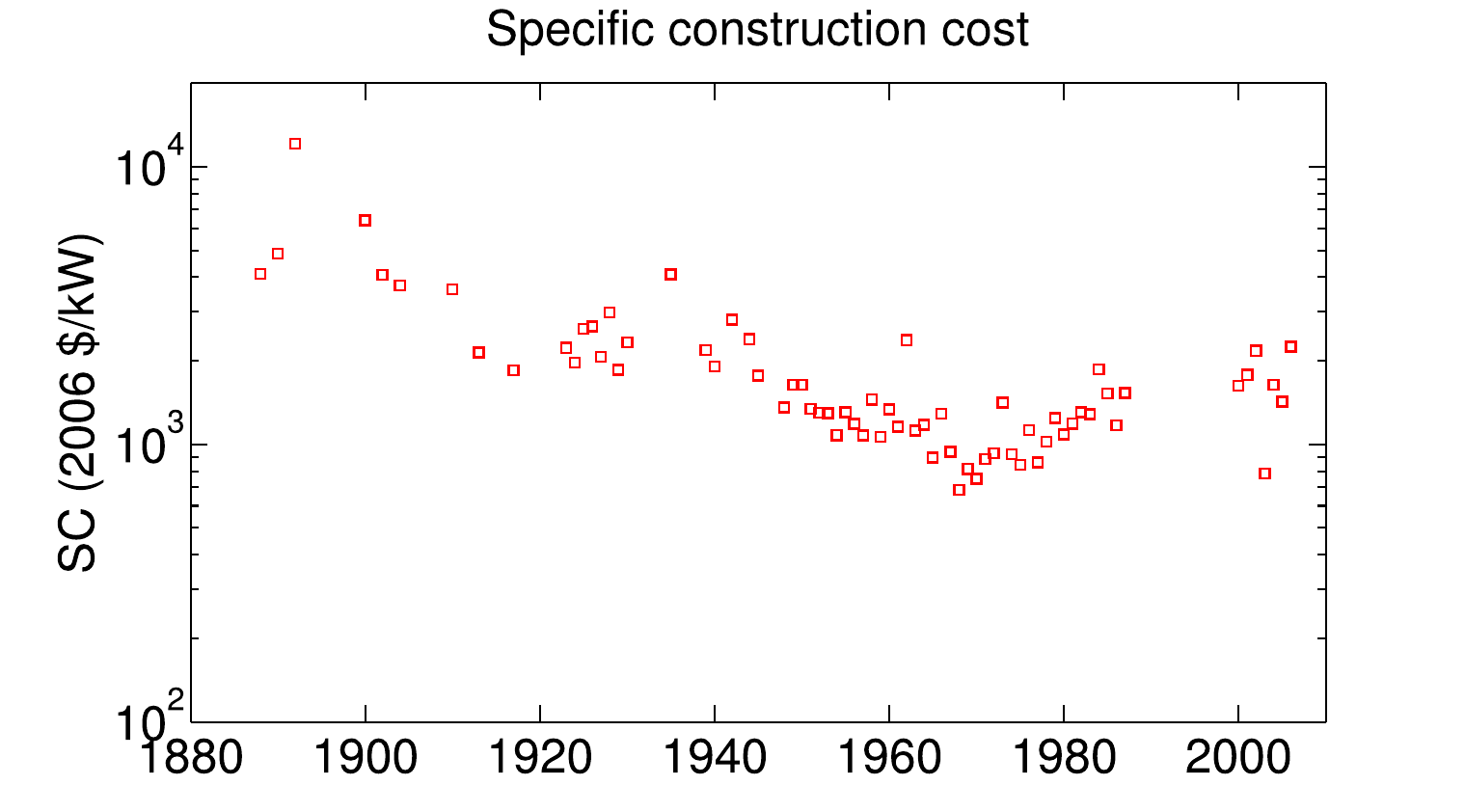}
\caption{Specific capital cost of coal plants. Plant costs generally decreased until 1970, when pollution regulation, increased construction times, and possibly other factors caused a rise. Source: \cite{EIA82,EIA87,NETL07} \label{spcc}}
\end{figure}

The specific construction cost (SC) is the cost of building a plant per kW of capacity installed. Specific construction costs decreased from the beginning of the industry until about 1970, then doubled between 1970 and 1987 (Fig. \ref{spcc}.) They appear to have been roughly flat since then, though we lack data for the period 1988-1999.

The specific construction cost is a leaf on our cost tree (a very important leaf), but could easily be the subject of a separate decomposition study unto itself, as Joskow \& Rose \cite{Joskow85} have done. Such a study would have a different scope from the present study, and involve collection of different data. Nevertheless, we can discuss the determinants of specific construction cost and in some cases provide numerical estimates of their impact. Four factors are frequently mentioned as important: economies of scale, add-on environmental controls, thermal efficiency, and construction inputs (both prices and quantities.) Note that

Economies of scale appear to have lowered construction costs as unit capacity grew (Fig. \ref{unit_number_size}).%
\footnote{A unit is a boiler plus turbogenerator. A plant may have one or more units.}
Joskow \& Rose study the effects of size on construction cost, and we combine their results with our size data to attempt a rough estimate of the cost reduction coming from unit capacity increases. They regress the real specific construction cost of U.S. coal units built between 1960 and 1980 onto a log linear function of unit capacity and other variables (such as the regional labor cost, the presence of pollution controls, and year dummies). In their model, SC depends on unit capacity $k(t)$ as $\text{SC} \propto  k(t)^a$. Under the simplest specification of their model, they find $a=-.183$. To estimate the size effect, we calculate the ratio of the size factor between two years, $t_1$ and $t_2$: $k(t_2)^a / k(t_1)^a$. This is the factor by which specific construction costs would have changed due to capacity changes, all else being equal. Results are shown for 3 periods in Table \ref{size_factors}.
\begin{table}[h]
\center
\begin{tabular}{lcc}
\hline
\textbf{Period} &$k(t_2)^a / k(t_1)^a$  &$\text{SC}(t_2)/\text{SC}(t_1)$\\
\hline
1908 -- 1970    &0.585                          &0.205\\
1970 -- 1989    &1.410                          &2.058\\
1989 -- 2006    &0.613                          &1.453\\
\hline
\end{tabular}
\caption{2nd column: Factor by which specific construction costs would have changed due to capacity changes, ceteris paribus, using Joskow \& Rose scaling exponent $a=-0.183$. 3rd column: Factor by which specific construction costs actually changed.}
\label{size_factors}
\end{table}

There are at least three limitations to calculating the size effect this way. One is that the estimate of $a$ is based upon units built between 1960 and 1980, and $a$ may be different at other times. Another is that measurements of scaling exponents depend sensitively on how samples of coal units are grouped; e.g. should one measure a single scaling coefficient for all coal units, or group them by pressure class, or by vintage, or by some other characteristic?%
\footnote{Joskow and Rose demonstrate this sensitivity using different multiple regression models, in which units were either all pooled together or grouped into 4 pressure classes. Among the various pressure classes and regression models, scaling coefficients varied from $-0.454$ to $+0.199$.}
The third limitation is more fundamental: it is not clear that specific construction cost actually has a capacity dependence of the form ${SC} \propto  k(t)^a$. Log-linear forms are common in the literature because they allow linear regression methods to be applied and because scaling phenomena often follow power laws. Nevertheless, no theory yet supports any particular functional form.

Increases in thermal efficiency can raise or lower the construction costs of plants \cite{MIT07}, by an amount that depends on the trade-off between higher materials and building costs but greater power. As engineering knowledge of a technology increases, it may be possible to obtain a higher thermal efficiency for the same materials and building costs. This results in lower construction costs (per unit power). However achieving a large jump in efficiency over a short period of time may lead to higher construction costs. Still, such a jump may still be attractive to investors if the plant is expected to have a high capacity factor and/or coal prices are high. The choice of thermal efficiency at a single point in time typically carries with it a tradeoff between higher efficiency/higher capital costs (where capital costs are influenced by the capacity factor) and lower efficiency/higher fuel costs.

At any given time in history, a range of efficiencies were accessible to the designers of plants and the utilities purchasing them. Utilities attempt to optimize the efficiency of the plants they purchase with respect to total costs, given their expectations about future fuel prices and the future load profile. The efficiency time series of Fig. \ref{efficiency_fuelcost} already accounts for this tradeoff, and depicts a change in the average believed-optimal efficiency over time.

Add-on environmental controls also raise plant costs, by an amount that may decrease over time with increased engineering knowledge. Joskow \& Rose estimate that sulfur scrubbers and cooling towers add about 15\% and 6\% respectively to the cost of plants. Besides raising plant costs directly by requiring new equipment, there is evidence that pollution controls also raised costs indirectly by increasing the complexity of the plant, which now required greater planning and longer construction times \cite{Joskow85,Cohen90}.

Finally, changes in the price or quantity of construction inputs will affect plant costs. Required quantities of materials and labor may decrease over time with increasing engineering knowledge, while changes in the rest of the economy may alter prices.

A long-standing trend of decreasing specific construction costs reversed direction around 1970 when costs began increasing, and the cause of this uptick has been a focus of interest. Several contributing factors have been identified. One is the introduction of pollution controls. Starting with the Clean Air Act in 1970, several laws were passed in the US requiring plants to add pollution controls to reduce the level of $\mbox{NO}_2$, $\mbox{SO}_x$, and particulates from flue gas emissions \cite{Joskow85,Gollop83}. The Clean Air Act of 1970 was followed by a number of similar acts between 1970 and 2003. While some utilities initially responded to pollution controls by switching to low-sulfur coals, others responded by installing de-sulfurization equipment known as sulfur scrubbers. Eventually all new plants were required to install scrubbers \cite{Gollop83}. This new equipment raised O\&M costs and decreased efficiency slightly, but mainly raised construction costs. 

Other causes cited for the increase in costs are inflation, interest rates, decreasing construction productivity and increasing construction times \cite{Joskow85,Cohen90}, reversed economies of scale (plants got smaller after 1970), and diminished opportunities and incentives to improve plant construction due to design variation and principal-agent problems \cite{McCabe96}. However, the relative contribution of each factor is unclear, and the cause of the uptick is only partially understood. \cite{Joskow85,McCabe96,Masters04}. Joskow \& Rose in particular find a large residual of unaccounted-for changes in cost after controlling for unit size, regional wages, utility experience, industry experience, pollution controls, indoor vs. outdoor construction, and ``first-unit effects.''%
\footnote{The first unit effect on a plant site is often more expensive than subsequent units because of one-time plant site costs.}
We refer the reader to several studies that examine more thoroughly the contribution of various factors \cite{Komanoff81,Gollop83,Joskow85,Hirsh89,Cohen90}.

A common representation of technological improvement is the experience curve or performance curve, a plot of the cost versus cumulative production of some item.  There is a large literature debating the value of experience curves as a model for technological improvement, which we review in Section~\ref{comm_v_tech}.
\begin{figure}[t]
\includegraphics[width=0.5\textwidth]{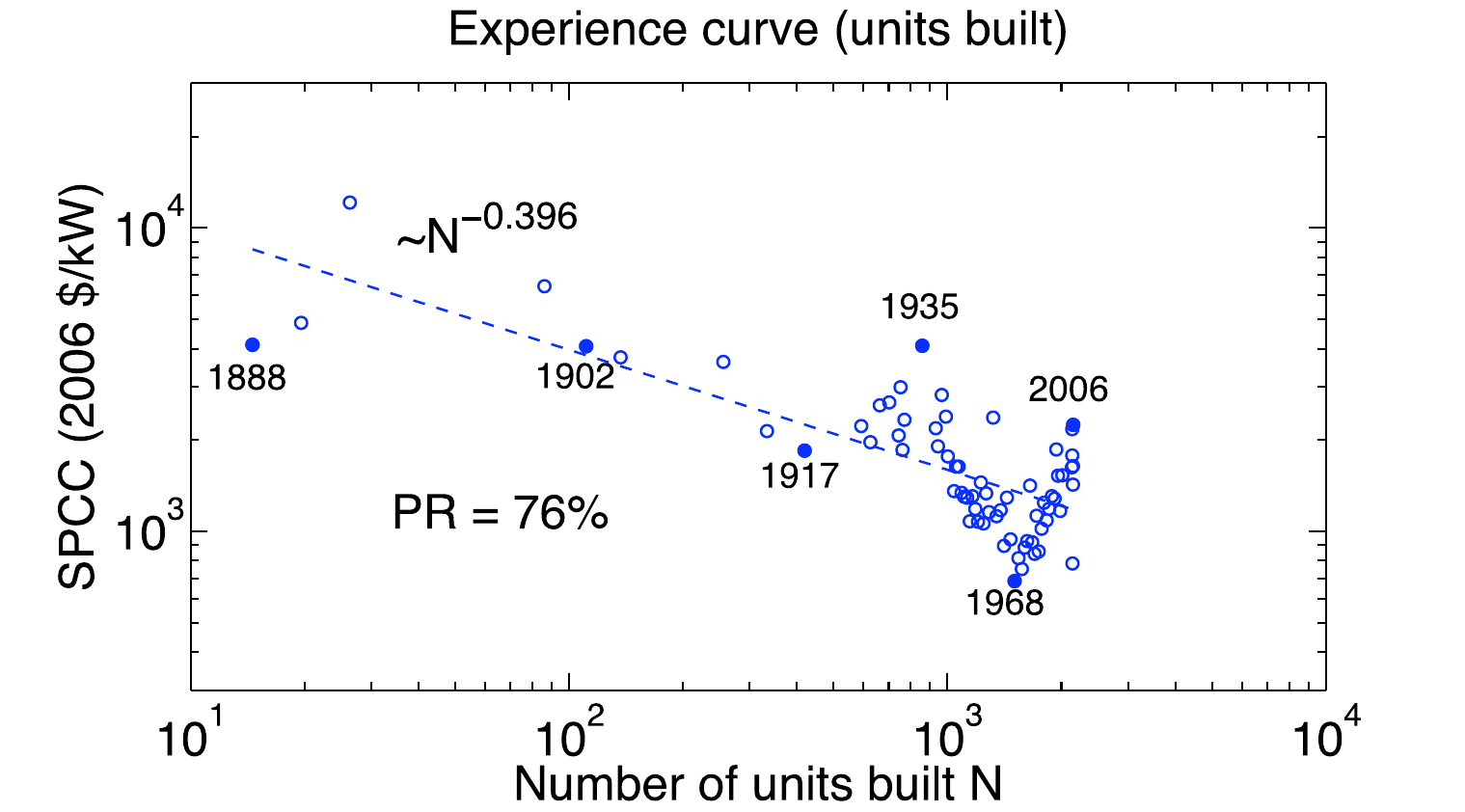}
\includegraphics[width=0.5\textwidth]{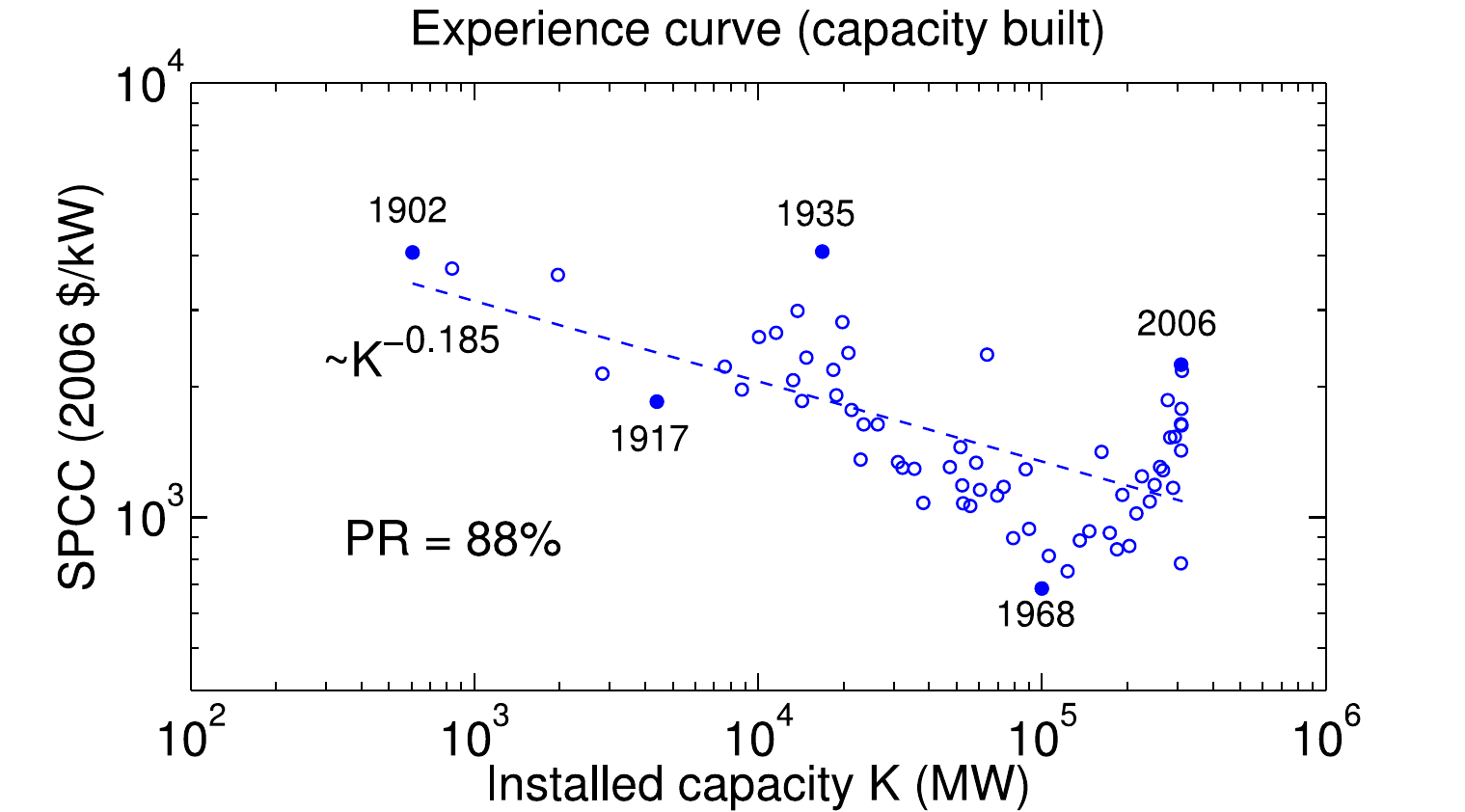}
\caption{Experience curves for plant construction cost, using two possible experience measures. Experience curves are a popular framework for predicting future costs of technologies. Source: \cite{EIA82,EIA87,NETL07,Platts,CensusBureau,AER06}}
\label{exp_curve}
\end{figure}
In Fig. \ref{exp_curve} we show the experience curve for plant construction costs. We use two different measures of experience: the number of coal units built, and the total capacity installed. Note that the cluster of points on the right side of Fig. \ref{spcc} is separated from the last point in 1987 by a gap due to missing data. However, in both plots of Fig. \ref{exp_curve}, this cluster of points joins up with those from the 1980s. This is consistent with experience curve hypothesis that costs are more correlated with changes in experience (as represented by cumulative production) than with time.

\begin{figure}[t]
\includegraphics[width=0.5\textwidth]{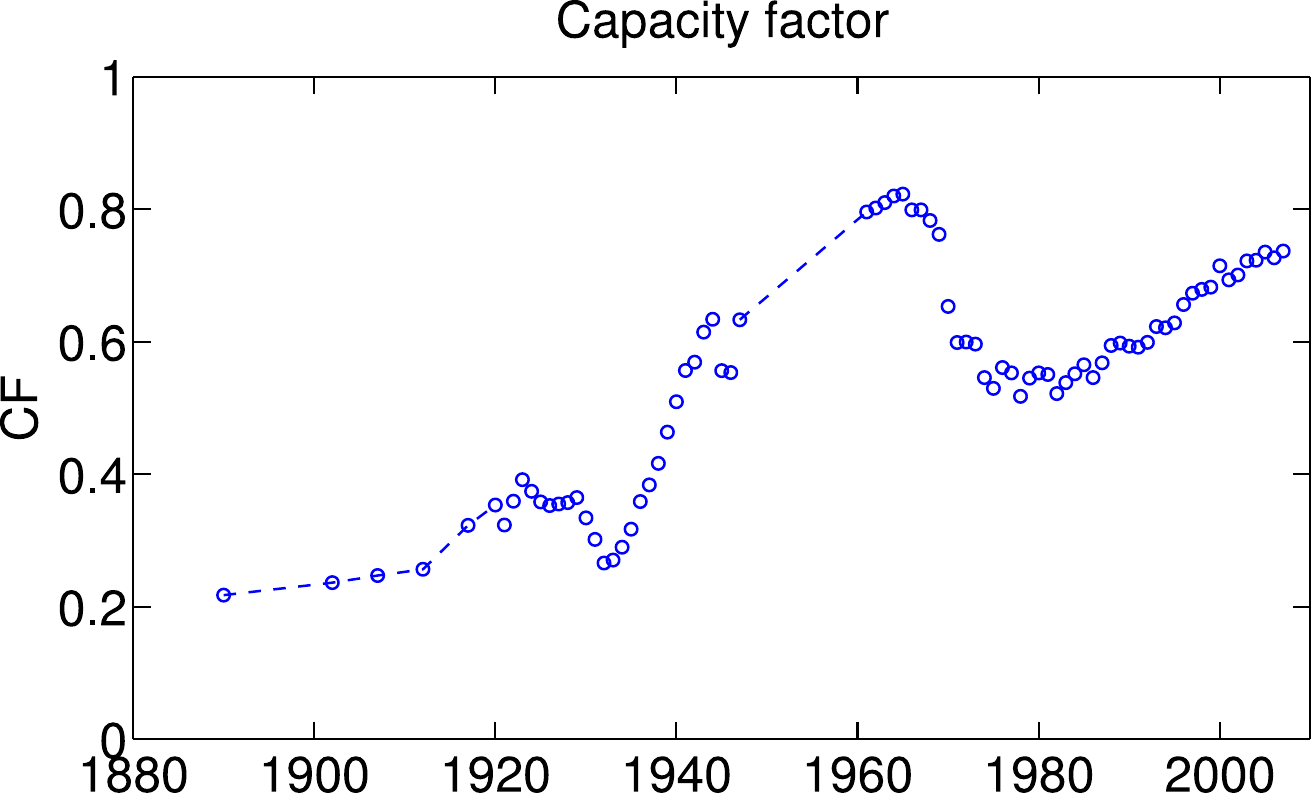}
\caption{Capacity factor of coal plants. Dashed lines are linear interpolations. Coal plants are typically base load plants, serving the persistent part of electricity demand at high capacity factor. The growth of the capacity factor, indicating greater utilization of the plant's capital, was a substantial factor reducing coal-fired generation costs. Source: \cite{CensusBureau,EEI92,Komanoff81,Platts,AER06}}
\label{cf}
\end{figure}

\subsubsection{Capacity factor}
The capacity factor (CF) is the amount of electricity produced divided by total potential production:
\begin{align*}
\text{CF}_t = \frac{\text{kWh production in year $t$}}{\text{kW capacity} \times 8760 \text{ hrs/year}}.
\end{align*}
The capacity factor measures the utilization of a plant's capacity, and is bounded between 0 and 1. 

Electricity generation incurs large fixed costs from plant construction, making high capacity factor -- high utilization of the plant's capital -- desirable to spread costs over the greatest possible production. The capacity factor is largely determined in advance, by the choice to build a base load plant or peaking plant. Demand for electricity varies hourly and seasonally. Building a coal plant with a high enough capacity to meet peak demand would leave the plant underused during periods of low demand, effectively raising capital costs. The cheapest way to meet varying electricity demand is with a combination of base load plants that have low operating cost (O\&M plus fuel) and run at high output rate to cover the persistent portion of electricity demand, and peaking plants that are relatively inexpensive to build and cover the excess portion of demand unmet by the base load plant.

Modern coal plants usually serve as baseload plants and therefore tend to have high capacity factors, around .7 -- .8 (Fig. \ref{cf}), though over the last 50 years the capacity factor has varied between .50 and .82. Capacity factors before 1940 were much lower. This may be because early plants required frequent maintenance and few devices existed that required electricity, resulting in less consistent demand for electricity throughout the day. The first major use of electricity was for lighting, particularly street lighting, which was only necessary a few hours each day. 

\subsubsection{Interest rate}
As mentioned earlier, we amortize the specific construction cost of plants (SC) to obtain the capital cost (CAP) using
\begin{align} \label{cap}
\text{CAP}_t = \frac{\text{SC}_t \times \text{CRF}(r_t,n)}{\text{CF}_t \times 8760 \text{ hours}}.
\end{align}
CRF is the capital recovery factor defined in Eq. \eqref{crf}. We take the plant lifetime $n$ to be 30 years for amortization purposes. This is both conventional and matches the longest bond maturities, the period over which capital payments would be made.%
\footnote{However, as a point of interest, it is quite possible that plant lifetimes are not 30 years, and moreover have changed with time. Longer plant lifetimes would effectively reduce electricity costs by spreading capital costs out over a longer period. One study found that most coal plants currently in use have lifetimes between 31 and 60 years, with an average of 49 \cite{Wynne}.} The resulting CRF for a given year was 1-2\% higher than the annual interest rate for that year.

\begin{figure}[t]
\includegraphics[width=0.5\textwidth]{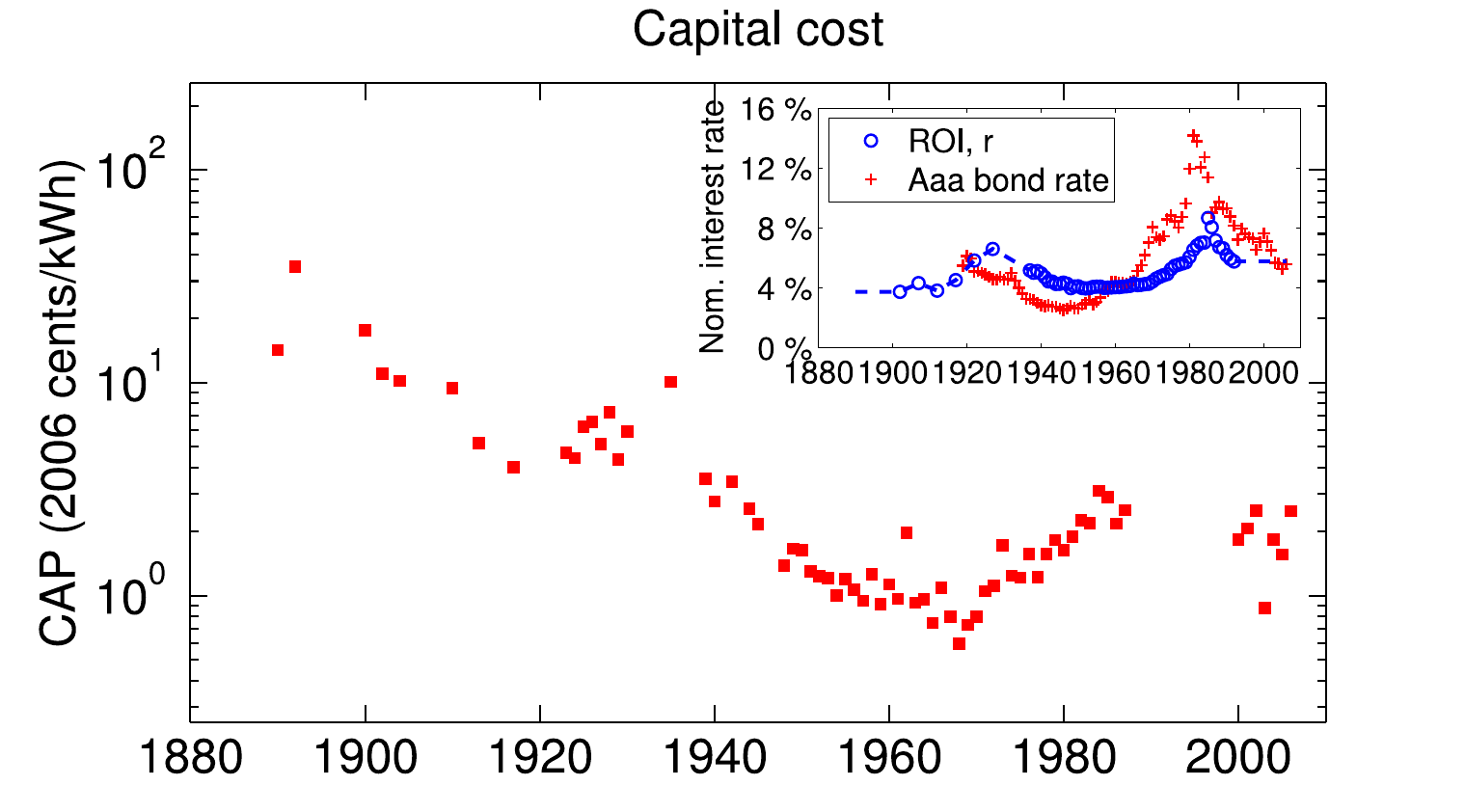}
\caption{The capital cost of coal plants. The capital cost accounts for the specific construction cost (Fig. \ref{spcc}), the capacity factor (Fig. \ref{cf}), and the interest rate (inset.) This series resembles that of the specific construction costs but is steeper due to increases in capacity factor. The inset shows the historical return-on-investment of electric utilities, used as the effective interest rate for amortizing the construction cost. The Aaa corporate bond rate is shown for comparison. Source: \cite{Fed,CensusBureau,FPC73,EEI92} \label{amcc_roi}}
\end{figure}

The effective interest rate $r$ used to calculate the capital recovery factor was the average return-on-investment (ROI) of the electric utility industry in each year (Fig. \ref{amcc_roi} inset.) The ROI is defined as the sum of annual interest and dividend payments made to investors divided by the industry's gross plant value. (Note that $r$ is a nominal interest rate.) This data was gathered from combined income statements and balance sheets for the electric utility industry as a whole \cite{CensusBureau,FPC73,EEI92}. The data is therefore not coal-specific, but we do not expect that interest rates charged to coal utilities would differ significantly from the average rate charged to utilities as a whole. The ROI varied between 4 and 8\%.

Interest rates had an important but transient effect; they contributed to the rise in costs seen during the 1970s and 1980s.

\subsubsection{Capital cost}
The capital cost (CAP) given by Eq. \ref{cap} combines construction costs, plant usage, and interest rates to obtain the contribution of capital to total generation costs (Fig. \ref{amcc_roi} main axes.) Its shape resembles that of the specific construction cost of Fig. \ref{spcc}. The most important factors reducing capital costs were decreasing construction costs and increasing plant usage.

\subsection{Total cost \label{totalcost}}
\begin{figure}[t]
\includegraphics[width=0.5\textwidth]{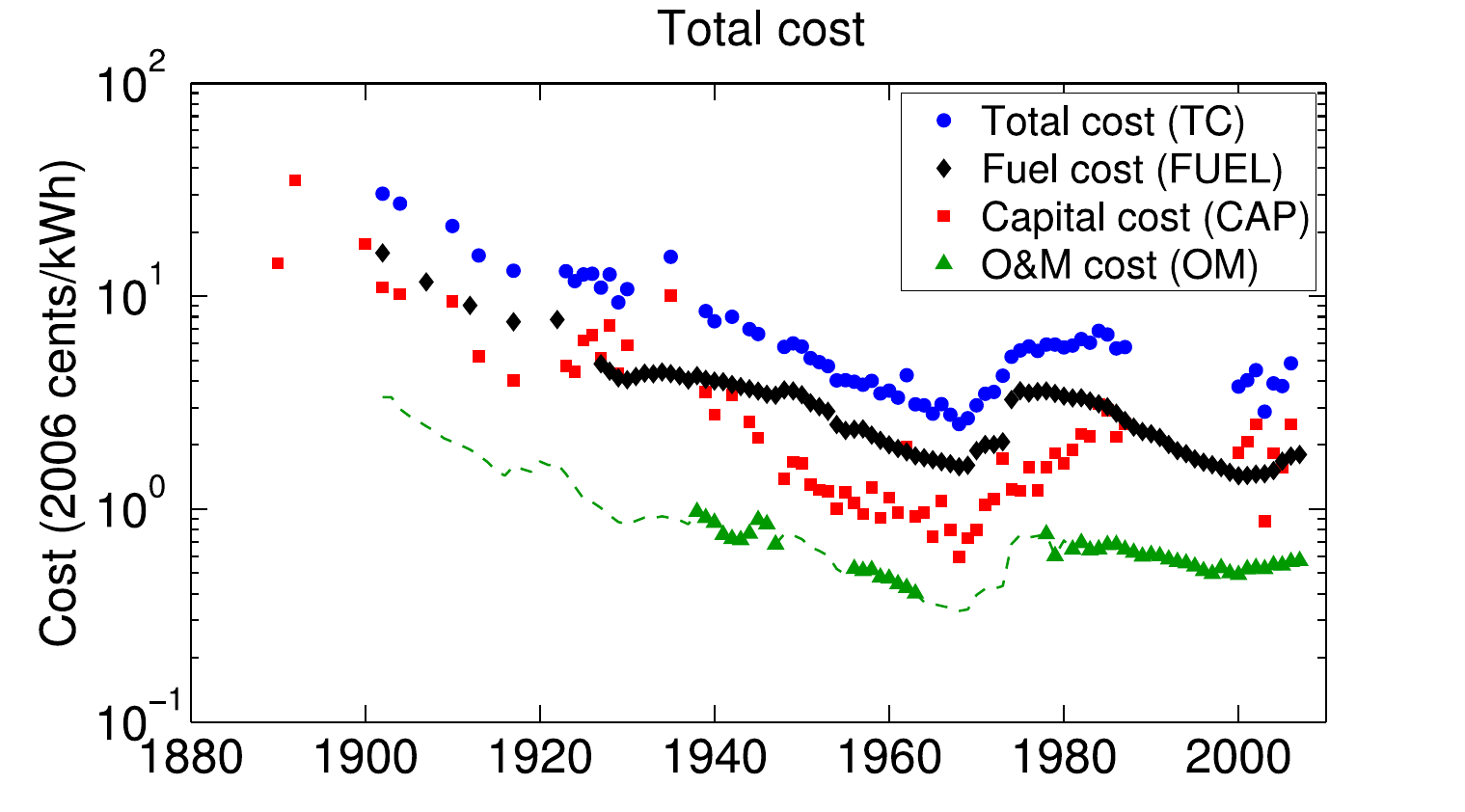}
\caption{The total cost of coal-fired electricity. The top curve is the sum of the fuel, capital, and O\&M curves below it. \label{tc}}
\end{figure}

\begin{figure}[t]
\includegraphics[width=0.5\textwidth]{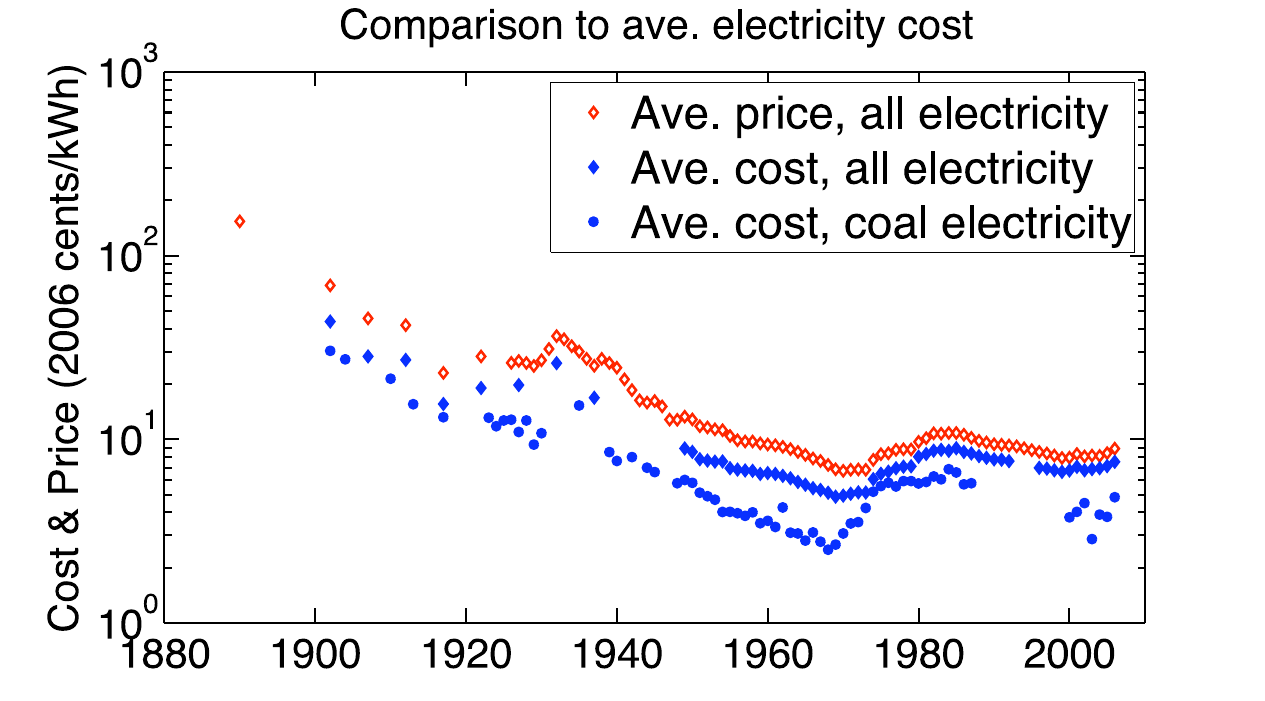}
\caption{The total cost of coal-fired electricity shown in Fig. \ref{tc}, along with the average price and cost of electricity from all generating sources. The cost from all sources, which largely consists of coal-fired generation, was used as a point of comparison to validate our built-up cost for coal-fired electricity. The former is expected to be somewhat higher because it includes electricity from more expensive sources. \label{price_comparison}}
\end{figure}

Figure \ref{tc} shows the total generation cost of coal-fired electricity (TC), along with the three major cost components. To check that the cost history constructed was approximately correct, we sought independent data to validate this series. To our knowledge, neither cost nor price data for coal-fired electricity exists, so we use the average price of electricity from all types of generation. Coal provided about half of all annual electricity production for the US throughout its history, so we expect the average price to be heavily influenced by the price of coal-fired electricity. In addition, we use historical data for transmission and distribution losses, taxes, and retained earnings (i.e. ``post-cost'' adjustments) to estimate the average \emph{cost} of electricity to obtain a theoretically closer series for comparison. The all-sources cost and price series described above, along with the reconstructed coal-fired cost, are shown in Fig. \ref{price_comparison}. These series compare favorably and validate the decomposition.

\begin{figure}[b]
\includegraphics[width=0.5\textwidth]{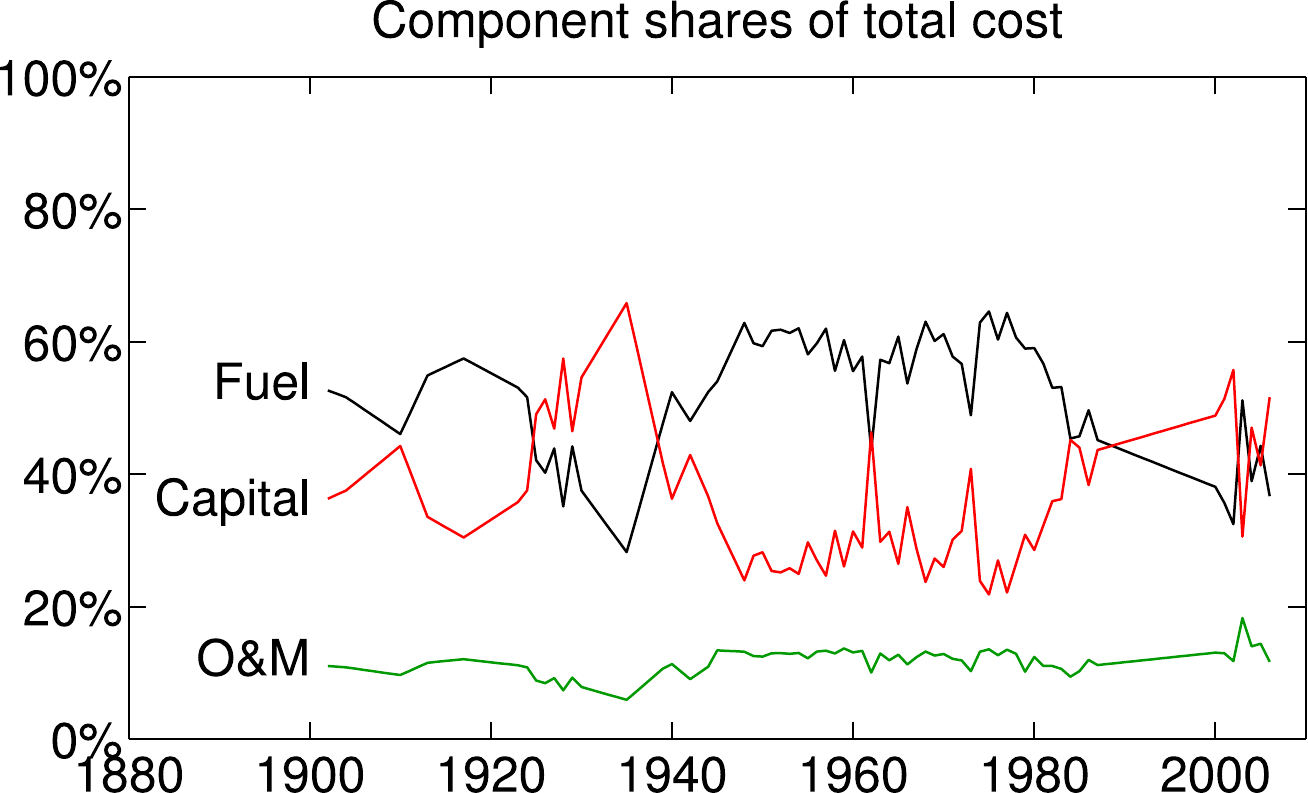}
\caption{The share of total generation costs contributed by each component. Fuel and capital have traded off in dominance of total generation costs over time, and are currently close in size. \label{component_shares}}
\end{figure}

Between 1970 and 1985, costs increased due to a number of unrelated developments that 
happened nearly simultaneously. Coal prices increased from high oil prices, anticipation of strikes, and increasing wages; O\&M costs increased because of added pollution controls; and plant construction costs increased from added pollution controls, diminishing productivity in the construction sector, and high inflation-driven interest rates.

In the next three sections we analyze the data presented above. In Section \ref{trends} we look at the long term cost trends of each variable. In Section \ref{variation} we look at the short term variation caused by each variable.

\begin{table*}[t]
\begin{center}
\caption{Decomposition of the change in cost of coal-fired electricity. The first two columns indicate the dollar amount of cost changes contributed by the individual variables of equation \eqref{main_equation}.  The last two columns indicate what percentage of the change each variable is responsible for. Negative percent contributions indicate that a variable opposed the change that was actually realized.}
\newcolumntype{.}{D{.}{.}{2}}
\begin{tabular*}{.80\textwidth}{@{\extracolsep{\fill}}l  .l.l  rr}
\hline															
\textbf{Variable $i$}	&	\multicolumn{4}{c}{\!\!\!\!\!\!\!\!\textbf{Effect on generation cost, $\Delta \text{TC}_i$}}							&	\multicolumn{2}{c}{\textbf{\% of $\Delta \text{TC}$ caused by $i$}}					 \vspace{6pt} \\
															
	&	\multicolumn{2}{l}{\!\!\!1902 -- 1970}			&	\multicolumn{2}{l}{\!\!\!1970 -- 2006}			&	\multicolumn{1}{r}{1902 -- 1970}		&	\multicolumn{1}{r}{1970 -- 2006}		\\
\hline															
TC	&	-27.24^a	&	\text{\!\!\!\!\!\textcent/kWh}	&	1.764^b&	\text{\!\!\!\!\!\textcent/kWh}		&	100.0	\%	&	100.0	\%	\\
\hline															
OM	&	-2.94	&		&	0.171	&		&	10.8	\%	&	9.7	\%	\\
FUEL	&	-14.08	&		&	-0.104	&		&	51.7	\%	&	--5.9	\%	\\
CAP	&	-10.22	&		&	1.697	&		&	37.5	\%	&	96.2	\%	\\
\hline															
OM	&	-2.94	&		&	0.171	&		&	10.8	\%	&	9.7	\%	\\
COAL	&	0.65	&		&	-0.088	&		&	--2.4	\%	&	--5.0	\%	\\
TRANS	&	-0.71	&		&	-0.286	&		&	2.6	\%	&	--16.2	\%	\\
$\eta$	&	-15.04	&		&	0.021	&		&	55.2	\%	&	1.2	\%	\\
$\rho$	&	1.01	&		&	0.249	&		&	--3.7	\%	&	14.1	\%	\\
SC	&	-3.38	&		&	1.544	&		&	12.4	\%	&	87.5	\%	\\
$r$	&	0.20	&		&	0.235	&		&	--0.8	\%	&	13.3	\%	\\
CF	&	-7.03	&		&	-0.081	&		&	25.8	\%	&	--4.6	\%	\\
\hline															
\multicolumn{7}{l}{$^a$ This is a 90\% drop from the 1902 generation cost.}\\															
\multicolumn{7}{l}{$^b$ This is a 57\% rise from the 1970 generation cost.}															
\end{tabular*}
\label{costchange_decomposition}
\end{center}
\end{table*}

\section{Analysis of cost trends}\label{trends}
We analyze the trends in total generation costs in two ways. First, we show the contribution of each major cost component to total generation costs (TC). Second, we show the contribution of each variable to changes in the total generation cost.

The major contributors to total generation cost are the fuel and capital components. The contribution from fuel has usually been 50-65\%, and that from capital 25-40\%. However, during two periods this ordering failed -- from about 1925 to 1939, when the roles of fuel and capital swapped places, and from 1984 to the present, where they contribute about evenly. The contribution of O\&M costs appears to have been relatively steady around 5-15\% during the whole history (Fig. \ref{component_shares}), though as noted earlier O\&M costs before 1938 are not based on data but inferred from fuel costs (Fig. \ref{om}). This estimated breakdown is similar to ones given in other sources \cite{Komanoff81,Cohen90}.

The second way we break down total costs is to decompose the \emph{change} in total cost contributed by each variable. That is, we calculate the change $\Delta \text{TC}_i$ in the total generation cost caused by each variable $i$. These contributions from individual variables sum up to the total change $\Delta \text{TC}$:
\begin{align*}
\Delta \text{TC} &= \Delta \text{TC}_\text{OM} + \Delta \text{TC}_\text{FUEL} + \Delta \text{TC}_\text{CAP}\\
&= \Delta \text{TC}_\text{OM} + \Delta \text{TC}_\text{COAL} + \Delta \text{TC}_\text{TRANS} + \Delta \text{TC}_{\eta}\\
& \quad + \Delta \text{TC}_{\rho} + \Delta \text{TC}_\text{SC} + \Delta \text{TC}_r + \Delta \text{TC}_\text{CF}\\
&= \sum_i \Delta \text{TC}_i
\end{align*}
The method for calculating each $\Delta \text{TC}_i$ is a generalization of Nemet's method \cite{Nemet06}. Our generalization is based on partial derivatives and is described in Appendix A.

Table \ref{costchange_decomposition} displays the results of this second decomposition during two periods, 1902--1970 and 1970--2006. The change in the generation cost $\Delta \text{TC}$ is given in the first row of the table; e.g. the generation cost dropped 27 \textcent/kWh between 1902 and 1970 and increased 1.8 \textcent/kWh between 1970 and 2006. The next three rows indicate the contributions to this change coming from the three major cost components: $\Delta \text{TC}_\text{OM}$, $\Delta \text{TC}_\text{FUEL}$, and $\Delta \text{TC}_\text{CAP}$. These contributions sum up to $\Delta \text{TC}$. The next eight rows decompose these contributions still further. Together these eight contributions also sum up to the total change in generation cost. Appropriate combinations of them will also sum to equal the contributions from OM, FUEL, and CAP. 

The last two columns of the table give the results in percentage terms. The percent change in generation cost effected by variable $i$ between years $t_1$ and $t_2$ is
\begin{align} \label{change_from_i}
\text{\% of change caused by $i$} = 100 \times \frac{\Delta \text{TC}_i(t_1,t_2)}{\Delta \text{TC}(t_1,t_2)}
\end{align}
where $\Delta \text{TC}_i(t_1,t_2)$ denotes the change caused by variable $i$ between years $t_1$ and $t_2$. By construction all percent contributions sum to 100\%. An implication of Eq. \eqref{change_from_i} is that variables which \emph{oppose} a given change in generation cost appropriately have negative percent contributions. For example, while in the first period total generation costs \emph{dropped} about 27 \textcent/kWh (representing 100\% of the total change), coal prices nevertheless increased slightly, and by themselves would have \emph{increased} generation costs by .65 \textcent/kWh (thus representing --2.4 \% of the total change.)

During the first period, two factors stand out most: increasing capacity factor, responsible for 26\% of the decrease in costs, and improving efficiency, responsible for 55\%. Following distantly are the specific capital cost (12\%) and O\&M cost (11\%). The fuel component as a whole was responsible for 52\% of the decrease in cost, and the capital component for 38\%.

Surprisingly, the construction cost was only responsible for 12\% of the change, despite decreasing by a factor of 5.4 during this time. However, it is important to realize that the magnitude of individual changes cannot be considered independently of other changes. If other cost-decreasing changes had not occurred simultaneously, the contribution of specific capital cost would have been much greater. Equally, if construction costs had not come down, it would have held up progress of the technology and changes in other factors would have been less important. Note that this is not a bug of the decomposition method used, but a fact occurring for any decomposition because changes in variables cannot be considered independently of other variables.%
\footnote{Consider for example, some quantity $y$ which depends on three variables $x_1$, $x_2$, and $x_3$:
\begin{align*}
y(x_1,x_2,x_3) = x_1 x_2 + x_3
\end{align*}
Suppose that $x_2$ changes by 100\%, and $x_3$ changes by 5\%. Does this mean that $x_2$'s contribution to the change is greater? Not necessarily; if $x_1$ is sufficiently small, then the change in the \emph{product} $x_1 x_2$ may be tiny compared with the 5\% change in $x_3$. Thus, there are important ``cross effects'' of variables, which cannot be avoided and are a general fact of decomposing changes.
}

During the second period, increases in plant construction costs contributed dramatically to the \emph{increase} in generation cost. The capital component as a whole rose sufficiently by itself to double total costs. The O\&M cost made a much smaller contribution to the increase, while a small change in fuel costs mitigated the total change in generation costs somewhat.

So far, we have studied the long term evolution of each variable and its effect on the generation cost. In addition to long term trends, though, some variables show significant short term variation. In the next section, we examine the influence of this variation on the generation cost.

\section{Analysis of cost variation}\label{variation}
We now look at the influence of short time (10 year) variations on the generation cost. We can do this using the same decomposition technique used above. First, we calculate $\Delta \text{TC}_i$ for each variable $i$ between two given years $t_1$ and $t_2$, as before. Then, instead of comparing $\Delta \text{TC}_i$ to $\Delta \text{TC}$, we compare it to the value of $\text{TC}$ in year $t_1$:
\begin{align*}
\text{\% variation caused by $i$} = 100 \times \frac{\Delta \text{TC}_i(t_1,t_2)}{\text{TC}(t_1)}.
\end{align*}
This measures how much the change in $i$ actually raised the generation cost between years $t_1$ and $t_2$. Note that the previous section asked ``how much of the generation cost change $\Delta \text{TC}$ does $\Delta \text{TC}_i$ represent?,'' regardless of the actual size of $\Delta \text{TC}$. Now we are asking whether $\Delta \text{TC}_i$ actually changed the cost significantly.
\begin{figure}[t!]
\includegraphics[width=.5\textwidth]{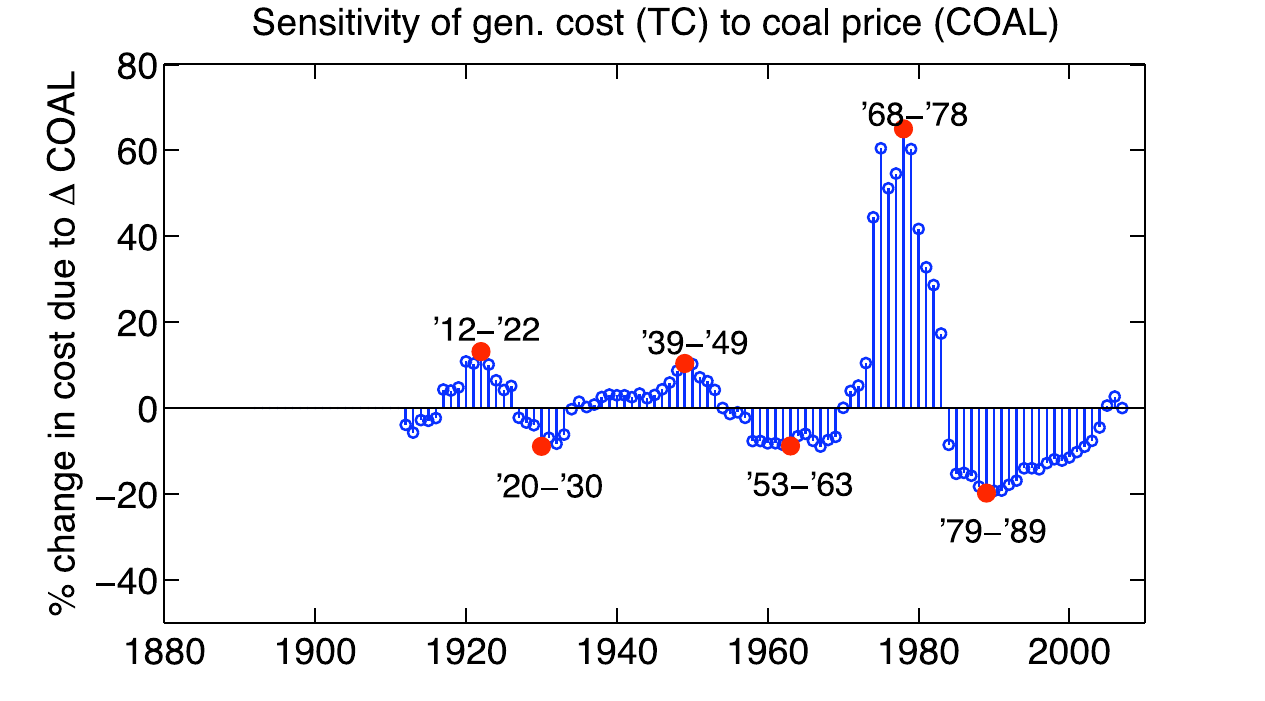}
\includegraphics[width=.5\textwidth]{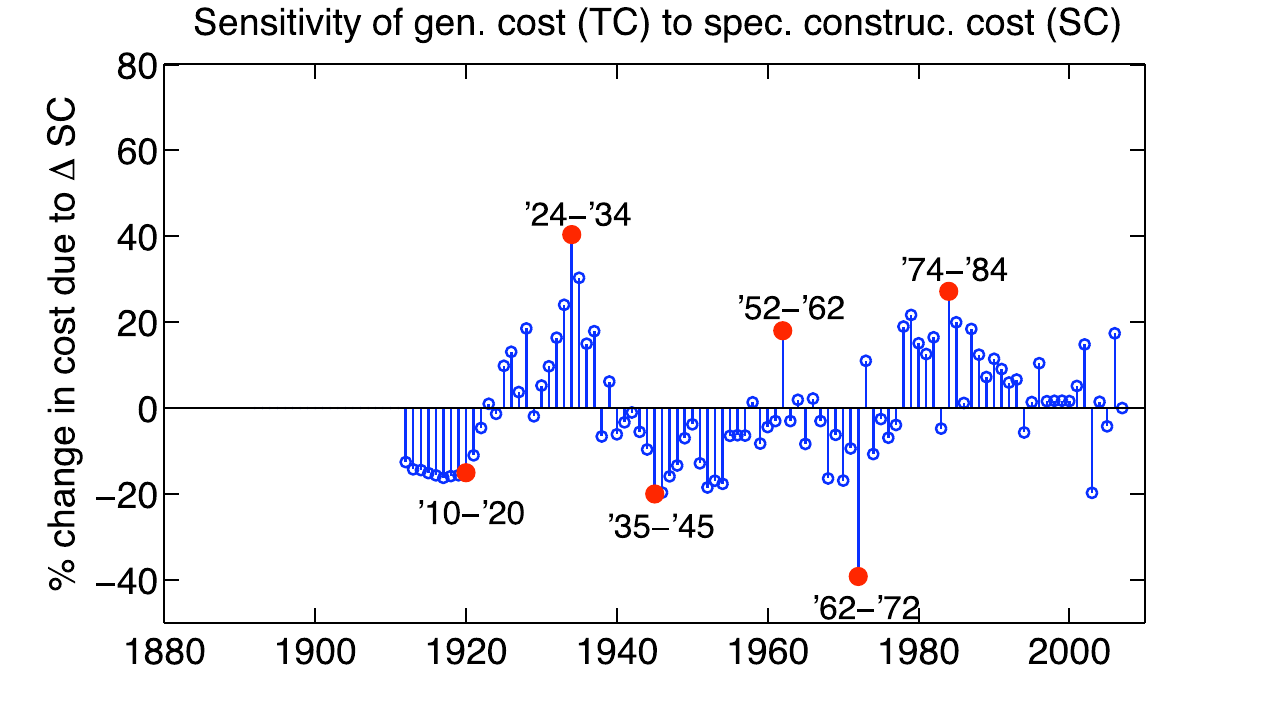}
\includegraphics[width=.5\textwidth]{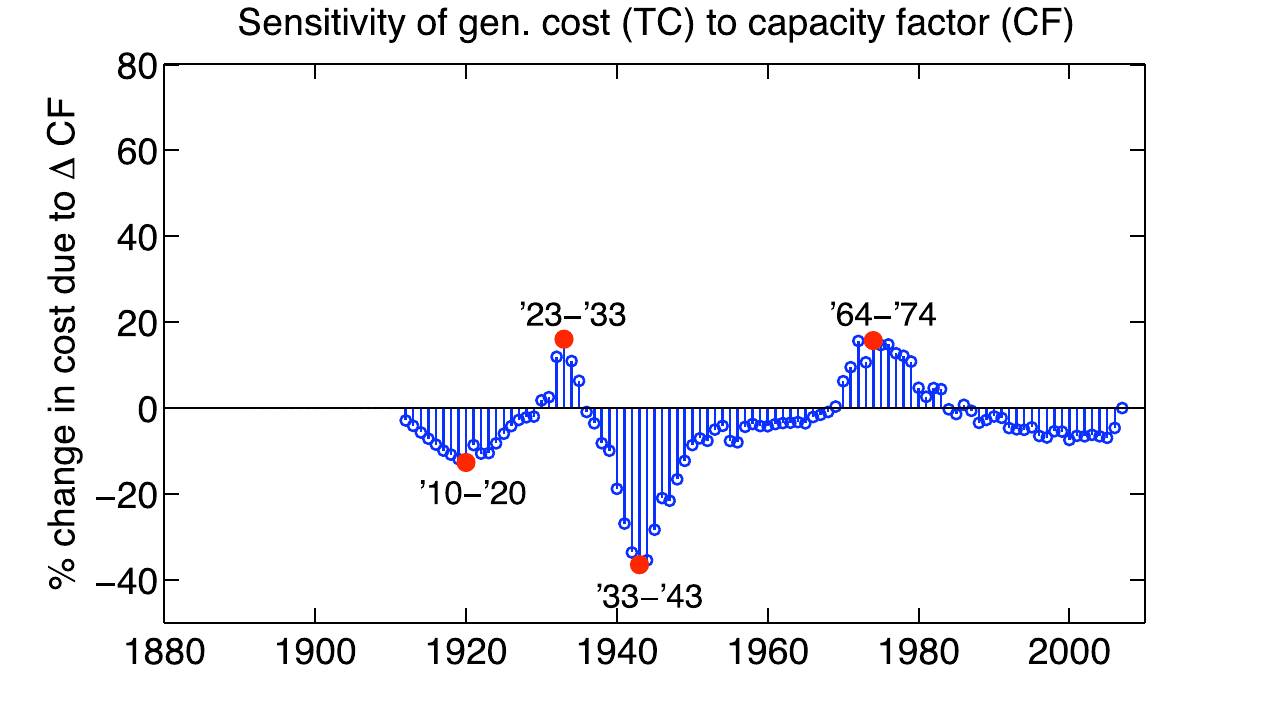}
\caption{The variation in total generation cost, TC, caused by changes in the coal price, COAL (top), specific construction cost, SC (middle), and capacity factor, CF (bottom). Heights of vertical lines estimate how large a change in the total generation cost was caused by the change in the coal price over the preceding 10 years. Time intervals with historically large changes are highlighted in red. These variables caused the largest cost changes out of those given in equation \eqref{main_equation}.}
\label{sensitivity}
\end{figure}

Rather than showing the variation in generation cost (TC) contributed by all the variables presented in the previous section, we focus on the three largest contributors of variation: the price of coal (COAL), the specific construction cost of plants (SC), and the capacity factor (CF). Fig. \ref{sensitivity} shows the influence these variables had on the generation cost. The height of the bar at each year $t$ shows the value of $\Delta \text{TC}_i(t-10,t) / \text{TC}(t-10)$; i.e. the size of the price change caused by variable $i$ over the previous 10 years. Thus, Fig. \ref{sensitivity} uses a sliding window which considers changes over every possible 10-year period.

The largest increase over any 10-year span came from coal price changes between 1968 and 1978,  when coal prices alone caused generation costs to increase by 64.3\%. The total increase in generation cost during this same period was 131.1\%. The remainder of the increase was driven mostly by increasing O\&M and capital costs.

Nevertheless, the construction cost and the capacity factor also contributed significant variation. Whereas coal price variation has tended to be smooth, with changes from consecutive years reinforcing each other, variation from construction costs is more noisy from year to year (although still following a long term trend.) Unlike coal, which is more or less the same product every year, each year's batch of new coal plants has unique characteristics that may make it more or less expensive than those of neighboring years. Thus, some of the spikes seen in the second plot of Fig. \ref{sensitivity} are due to two batches of plants 10 years apart that happen to have significantly different average costs.
\begin{figure}[t]
\includegraphics[width=.5\textwidth]{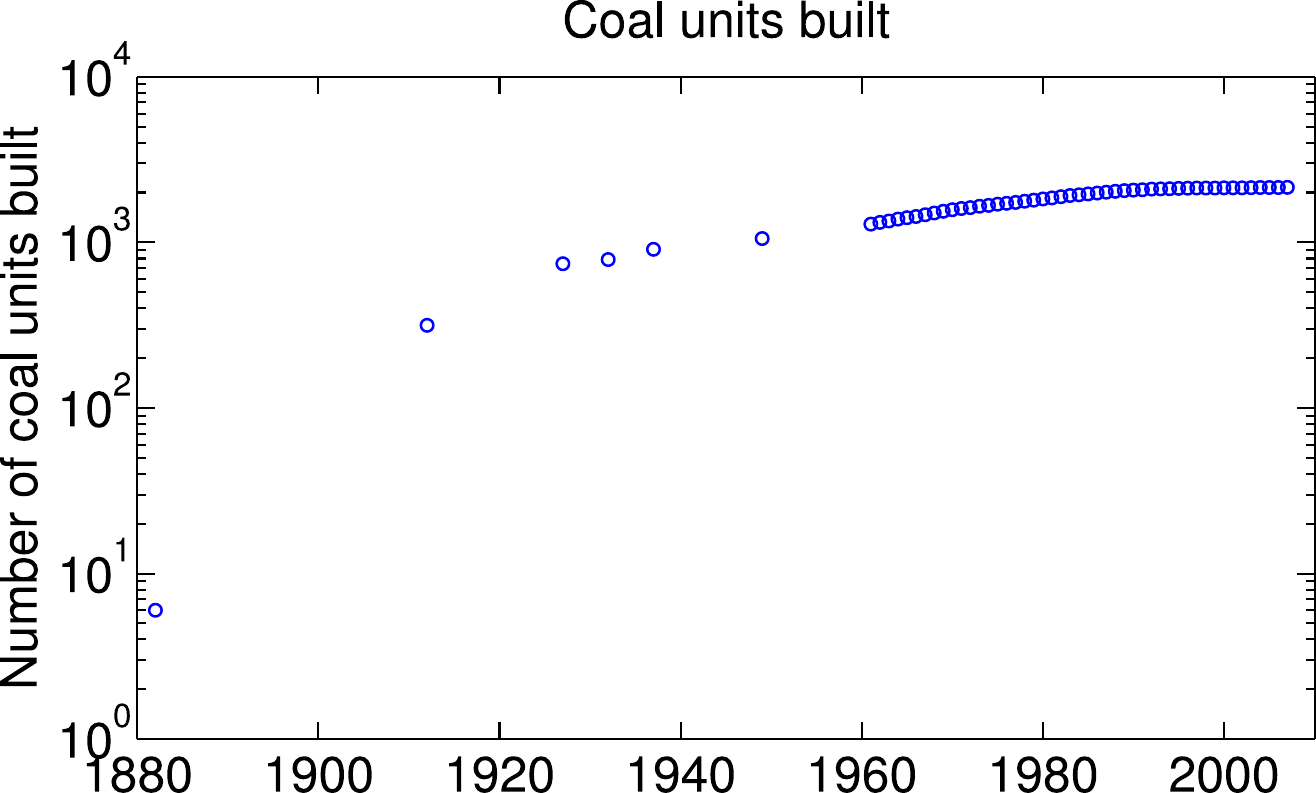}
\includegraphics[width=.5\textwidth]{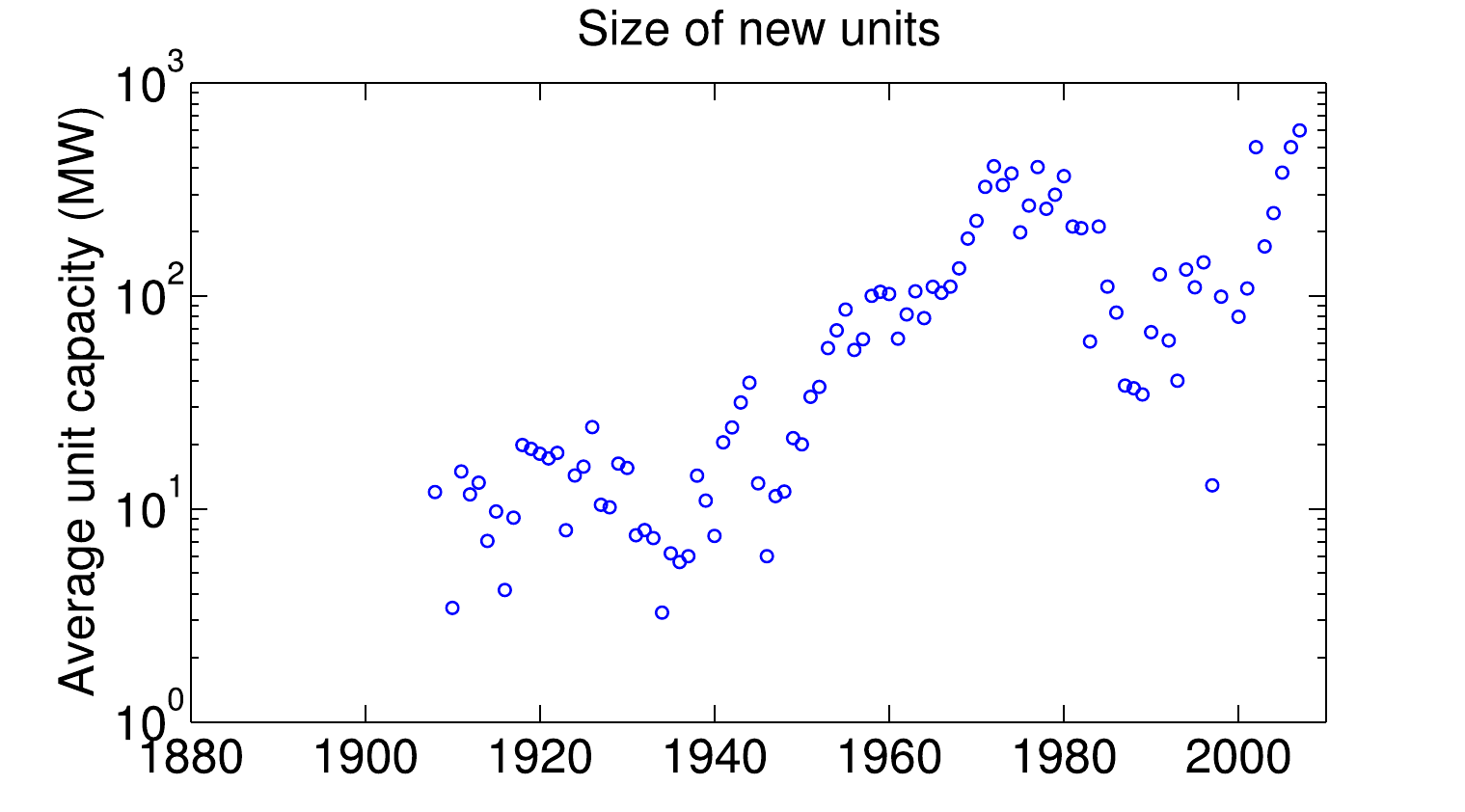}
\caption{Number of coal units produced (top), and geometric average capacity of new units (bottom). Source: \cite{Platts}}
\label{unit_number_size}
\end{figure}

Finally, although the capacity factor also contributed large variation, it is mostly an exogenous factor. It depends on industry-wide trends, rather than on any factors unique to coal-fired electricity, as mentioned before.

\section{Future Implications}\label{extrapolation}
The history of coal-fired electricity suggests there is a fluctuating floor to its future costs, which is determined by coal prices. In the following sections, we elaborate on this point. 

To motivate our discussion, we note that the driving variable behind the capital cost has been the specific construction cost, while the driving variable behind the fuel cost since efficiencies stopped improving has been the price of coal. The specific construction cost has tended to follow steady long term trends.  In contrast, the price of coal seems to vary randomly, with no clear trends.   As discussed in the next section, we hypothesize that this difference in price behavior is due to a fundamental distinction between a \emph{commodity} and a \emph{technology}. First, we suggest a framework for discussing these qualitative differences in price evolution, and then we apply the framework to the present example of coal-fired electricity.

\subsection{Commodities vs. pure technologies} \label{comm_v_tech}
In this section we present the conjecture that the prices of commodities and technologies evolve in fundamentally different ways. The meaning of these terms will be clarified momentarily, but the relevance to coal-generated electricity is that coal is essentially a commodity, whereas the construction cost of a plant is closer to (though not purely) a technology.

A commodity is a raw material or primary agricultural product that can be bought and sold. A standard assumption in the theory of finance is that markets are efficient; roughly speaking, this means it should not be possible to make consistent profits by arbitrage of commodities using simple strategies. If the price of coal were too predictable using simple methods, such methods would become common knowledge, and the buying and selling activity of speculators would affect prices in a way that would destroy their predictability. More specifically, one expects prices to follow a random walk to some approximation. Furthermore, the activity of speculators should bound the long-term rate of growth or decline of the price. Otherwise it would be possible to make unreasonable profits -- more reliably than could be made in alternative investments -- by either buying coal and hoarding it or by short selling coal.%
\footnote{Other ways to speculate on the price of coal are to buy or sell coal companies and to buy or sell land containing coal deposits.}

A pure technology is a body of knowledge, such as knowledge of a manufacturing process. We will say a product behaves like a pure technology when accumulated knowledge, reflected in changes to the technology's design, is a greater determinant of its cost evolution than speculation. For example, the fuel component of coal-fired plants was initially more pure technology-like. Changes to boiler and plant design \cite{BlackVeatch96} resulted in growth of thermal efficiency and pure technology-like cost evolution in the fuel component (which depends on the thermal efficiency.) Although the input cost of coal could and did change, its changes had less impact on the fuel component cost than these pure design changes. In the present day, with many of the efficiency improving design changes already exploited, changes to input costs determine changes in the fuel component to a greater extent, making the fuel component more commodity-like.%
\footnote{It's also reasonable to wonder the extent to which coal itself is a pure technology versus a commodity. This question is further complicated by the fact that coal is influenced by scarcity factors as well as technological factors. Anecdotal evidence makes it clear that both of these are influential. Both of these factors can, and probably did, cause changes in the cost/price of coal. (Although it is a bit interesting that they should stay so balanced with each other over a 130 year period. We say this noting that the dramatic rise in the 70s was largely due to exceptional circumstances having little to do with scarcity or technology factors.) Nevertheless, the theory of finance bounds the \emph{anticipated} long-term changes that can occur. This allows for unanticipated changes that may have any size. It also allows for anticipated long term changes that are sufficiently slow that they are not worth taking advantage of.}

Commodities and pure technologies are ends of a spectrum, and probably no real products are completely one or the other. The specific construction cost, for example, depends on construction technologies -- that knowledge of construction process and design which separates efficient uses of materials and labor input from inefficient uses. However, specific construction cost also depends on material commodities, such as steel. Thus specific construction cost is partly a technology and partly a commodity. Likewise coal is not completely a commodity, since knowledge of coal mining methods also affect its price, in addition to the speculation mechanisms mentioned above. A third example of this mixed composition is a photovoltaic system. The price of a photovoltaic (PV) system depends in part on commodities, such as metals and silicon. Increases in the price of silicon, for example, caused increases in the cost of PV cells in 2005-2006. Nonetheless, silicon is far from a raw material -- processing dominates the cost of mono and polycrystalline silicon, rather than commodity input costs. Thus, while the price of PV cells is partially commodity-driven, it may behave more like a pure technology than a commodity. Its historical behavior suggests this is true (see the silicon price time series in Nemet \cite{Nemet06}), though this possibility deserves further investigation.

\subsection{Modeling a commodity} \label{modeling_commodity}
The price of a commodity is usually modeled using time series methods. Time series models form a class of nested models that can be arbitrarily complex, but the simplest model and conventional starting point is an autoregressive model of order 1 (AR(1).)

The AR(1) model is defined by the equation
\begin{align*}
p_t = \gamma p_{t-1} + \mu + \epsilon_t.
\end{align*}
Here $p_t$ is the logarithm of the price in year $t$, $\epsilon_t$ is a random variable, and $\gamma$ and $\mu$ are parameters. The noise term $\epsilon_t$ is assumed to be uncorrelated in time and normally distributed with variance $\sigma_{\epsilon}^2$. The three parameters $\gamma$, $\mu$, and $\sigma_{\epsilon}$, combined with an initial condition $p(0)$, determine a given AR(1) process.%
\footnote{``Order one'' refers to the number of earlier prices the model regresses the current price onto. An AR(3) model, for example, would be specified as $p_t = \gamma_1 p_{t-1} + \gamma_2 p_{t-2} + \gamma_3 p_{t-3} + \mu + \epsilon_t$.}

The parameter $\gamma$ determines the long-term behavior of the process. When $\gamma=1$, the process is a random walk with drift, and when $\mu > 1$ the mean of the distribution of future prices grows without bound. When $\gamma<1$, the process is stationary and the distribution of future prices asymptotically has a fixed variance. The closer $\gamma$ is to 1, the larger the variance of the limiting distribution becomes. For $\gamma=1$, the parameter $\mu$ generates drift in the random walk, but for $\gamma < 1$ it sets a non-zero mean value around which the price moves.

To see how closely the coal price resembles a random walk, we fit the historical coal price to the AR(1) model using least squares regression. The parameter values obtained are $\gamma \approx .956 \pm .053$, $\mu \approx .144 \pm .174$, and $\sigma_{\epsilon} \approx .080$. Note that $\gamma$ has an error interval that includes $\gamma=1$.  Since $\gamma=1$ has qualitatively different behavior than $\gamma<1$, it is important to know the likelihood that the underlying process actually has $\gamma=1$ and is therefore a random walk. We apply various Dickey-Fuller tests of the null-hypothesis that the underlying process is a random walk, and find that in no test can this possibility be rejected.  Thus, our hypothesis is validated in two different ways:  First, to within statistical error $\gamma \approx 1$, suggesting a random walk, and second, to within statistical error the drift rate $\mu \approx 0$, suggesting no long term trend in coal prices.

\subsection{Modeling a technology}
To complete our discussion contrasting commodities and technologies, we describe the two most common models, experience curves (which are typically assumed to be power law) and extrapolation of time trends (which are typically assumed to be exponential).

The cost of a technology is often modeled in terms of an experience curve, a plot of its cost versus its cumulative production. The motivation for plotting cost against cumulative production often uses the following chain of reasoning: cumulative production measures ``experience'' with a technology; as a firm or industry gains experience it makes improvements to the technology, which cause cost reductions; therefore there may be a simple, predictable relationship between cumulative production and cost. Alternatively, one can argue that cumulative production is an indicator of profit-making potential, which drives the level of effort directed at improving a given technology \cite{Sinclair00}.

In fact, a simple relationship is frequently observed: empirical experience curves often appear to obey power laws. There is a large literature attesting to this regularity \cite{Dutton84,Thompson08}. Partly for this reason, the experience curve, combined with the assumption of a power law form, has become the prevailing method for extrapolating future costs \cite{Farmer07}. The power law functional form can also be derived from theoretical models \cite{Muth86,McNerney09}. 

We find it useful to distinguish the \emph{data} from any particular \emph{hypothesis} about its shape. We therefore use the term experience curve to mean the plot of cost versus cumulative production, whatever its shape. We reserve the term Wright's law for the hypothesis that an experience curve should follow a power law. While we acknowledge that many researchers implicitly include the power law functional form with the definition of an experience curve, we prefer to refer to the data in a theory-neutral way.

Alternatively, the cost of technologies has been modeled as decreasing exponentially with time; this evolution can be viewed as a generalization of Moore's law.\footnote{See Nagy et al. \cite{Nagy09}. Moore's law originally states that the number of transistors per integrated circuit doubles about every two years on average. This regularity can be restated in terms of the cost per transistor, which also decreases exponentially over time. The new statement of Moore's law has the advantage of being expressed in general terms comparable across technologies (i.e. cost versus time).} Moore's law (an \emph{exponential} decrease in cost with \emph{time}) and Wright's law (a \emph{power law} decrease in cost with \emph{cumulative production}) are not necessarily incompatible. They could simultaneously hold when cumulative production grows exponentially with time.\footnote{See, for example, Sahal \cite{Sahal81}.}

Two kinds of criticisms have been brought against the program of plotting cost data and fitting it to curves.\footnote{These criticisms have often been directed against experience curves in particular, but could apply to any low-dimensional model of cost evolution.} One is that it buries too much detail about the processes driving cost reductions \cite{Sinclair00}. Another criticism questions the predictive power of different functional forms, since they have not been rigorously compared against each other. Further progress in this area requires both careful comparison of the predictive ability of different effective models and study of the causes underlying observed relationships. Nagy et al. have recently compared the historical prediction accuracy of various functional forms proposed by researchers to describe cost evolution \cite{Nagy09}, including Wright's law and Moore's law. 

We note that any mathematical description of a technology's cost evolution is likely to fail when the technology itself substantially changes. A famous example is the Model T Ford, which dropped smoothly in cost from its introduction in 1909, when Henry Ford announced he would make a car that the common could afford, to 1929, when he ceased production. During this period the cost of Model T's follow Wright's law quite well \cite{Abernathy74}. After that Ford produced other models with better performance and, not surprisingly, higher costs. This is an important point; in general an extrapolation may only hold when the product faces fixed performance criteria. When a product is called upon to meet criteria it did not previously meet, such as lower pollution or higher safety, its cost evolution may show discontinuous behavior.

The two models presented are the most common models for what we call technology-like cost evolution. The important property they share is that both involve predictable, long-term, decreasing costs. With this in mind, we now return to the specific case of coal-fired electricity.

\subsection{Application to coal-fired electricity}
Plant construction costs have not dropped significantly in 40 years. Pollution controls have redefined coal plants to such a degree that they are not really the same product as that produced in the first 80 years. But let us imagine what would happen if coal plant construction costs were to revert to the same technology-like cost evolution they followed during the first 80 years. Plant costs would drop again, and dropping plant costs would cause fuel -- undergoing commodity-like evolution -- to become the dominant cost. The dominance of fuel costs would in turn lead to greater sensitivity of generation costs to fluctuations in the price of coal. Thus, the price of coal would determine a fluctuating floor on coal-fired generation costs.

This scenario assumes no dramatic change in thermal efficiency. As mentioned in Section \ref{efficiency}, efficiency in the U.S. has remained flat the last 50 years for several reasons, while it has increased elsewhere. However, it would take a substantial increase in efficiency to yield a substantial reduction in fuel costs --- of a magnitude comparable to historical reductions. For example, if the average efficiency increased from 33\% to 43\%, the fuel cost would be decreased by a factor of $(1/\eta_2)/(1/\eta_1) = (1/.43)/(1/.33) = 0.77$, a far cry from the (physically unrepeatable) factor-of-10 improvement from the start off the industry.

For any plant design, once efficiency improvements have been exhausted, the price of coal will set a floor on total costs. If we consider the possibility of carbon capture and storage--- which is likely to increase capital costs initially---the potential reduction in total costs over time would similarly be limited by the fluctuating floor defined by the coal price.

\section{Conclusions}
We make several methodological advancements in this paper. We consider total generation costs (decomposed into components) rather than costs of single components alone, use data over a relatively long time span (over a century and covering the lifetime of the industry), and use a physically accurate bottom-up model of costs.  As part of this decomposition, we model coal prices as a random walk while modeling construction costs and O\&M costs as an improving technology.

We find evidence that the fuel and capital costs of coal-fired electricity evolve differently due to different behaviors in coal prices and plant construction costs. Coal prices have fluctuated with no trend up or down, while plant construction costs have followed long-term trends. The behavior of coal prices is consistent with the facts that coal is a freely traded commodity. Plant construction cost may follow a pattern of long term reduction because it is only weakly influenced by these effects and is able to realize improvements typically seen in technologies over time.

Such a difference in the behavior of coal prices and plant construction costs would yield different behavior for the fuel and capital cost components. Although historically both the fuel and capital components improved at similar rates, the main driver of improvement in fuel costs -- thermal efficiency -- has been unchanged since the early 1960s. Without a substantial improvement in the average thermal efficiency, the main driver of change in fuel costs would be the coal price.  Coal prices, in contrast, are statistically neither decreasing nor increasing, and so provide a statistically fluctuating floor on the overall generation cost, with no clear long-term trend.

Although all the analysis in this paper is specific to coal, we hypothesize that a similar analysis might apply to other fossil fuel-based sources of electricity, such as natural gas.  In fact, because natural gas and oil are both traded on exchanges, with standardized futures contracts, it is easier to speculate in them than it is in coal.  Thus we would expect that the commodity part of our model will apply to them even more strongly. This would suggest that the cost of the major fossil fuel-based sources of electricity are all constrained by a noisy floor determined by fuel prices.

\section{Acknowledgments}
We gratefully acknowledge financial support from NSF Grant SBE0738187. All conclusions presented here are those of the authors and do not necessarily reflect those of the NSF. We are grateful to Margaret Alexander and Tim Taylor for substantial help obtaining source materials. We thank Arnulf Gr\"ubler for helpful conversations and suggestions. We thank Charlie Wilson for several valuable conversations and extensive feedback on an early draft, and Dan Schrag for a challenge that led to the idea for this paper. We also thank anonymous reviewers for their insightful comments.

\appendix
\section{Change decomposition}
Consider the following problem. We have a function $f=f(x,y)$, and during some period of time $f$ changes as a result of simultaneous changes to $x$ and $y$. We want to know how much of the change to $f$ each variable is ``responsible for.''

To be more precise, let $\Delta f$ be the change in $f$. We would like to decompose $\Delta f$ into 2 terms, corresponding to the change contributed by each variable:
\begin{align*}
\Delta f = \Delta f_x + \Delta f_y,
\end{align*}
where $\Delta f_i$ denotes the change in $f$ resulting from the change in $i$. A way to do this decomposition is suggested by the calculus identity
\begin{align*}
df = \frac{\partial f}{\partial x} dx + \frac{\partial f}{\partial y} dy.
\end{align*}
The trick is to generalize this expression to finite rather than infinitesimal changes. More to the point, we need to be able to take common combinations of variables -- e.g. products, quotients -- and express changes to these combinations in ways appropriate for finite differences.

For example, consider a product of two variables, $f(x,y) = xy$. The differential $f$ is
\begin{align*}
df = x\;dy + y\;dx
\end{align*}
suggesting the decomposition for finite differences is $\Delta f = x \Delta y + y \Delta x$. However, the correct expression is
\begin{align*}
\Delta f &= x \Delta y + y \Delta x + \Delta x \Delta y.
\end{align*}
In addition to the two expected terms, a third cross term containing both differences appears. As the differences become small, the cross term will vanish more quickly than the other terms, recovering the calculus limit $df = x\;dy + y\;dx$. However, for finite differences, the cross term potentially introduces a large residual if ignored.

In order to decompose the change in $f$ into just 2 pieces, we evenly split the cross term into the other terms:
\begin{align*}
\Delta f &= x \Delta y + \frac{1}{2}\Delta x \Delta y + y \Delta x + \frac{1}{2}\Delta x \Delta y\\
&= (x+\frac{1}{2} \Delta x) \Delta y + (y + \frac{1}{2}\Delta y) \Delta x
\end{align*}
The first term may be interpreted as the change in $f$ due to change in $y$; the second as the change in $f$ due to change in $x$. We therefore have our desired decomposition for the case of products of 2 variables:
\begin{align*}
\Delta f_x \equiv (y + \frac{1}{2}\Delta y) \Delta x\\
\Delta f_y \equiv (x + \frac{1}{2}\Delta x) \Delta y
\end{align*}
which by construction has the desired property $\Delta f = \Delta f_x + \Delta f_y$. Similar rules can be derived for products of 3 or more variables, for quotients, and other expressions.

\bibliographystyle{model2-names}

\end{document}